\documentclass[letterpaper, 11pt, final, onecolumn]{IEEEtran}
\usepackage[margin=1.0in]{geometry}
\usepackage{times}
\usepackage{cite}
\usepackage{url}
\usepackage[pdftex]{graphicx}
\usepackage{subfigure}
\usepackage{rotating}
\usepackage{rotfloat}
\usepackage{xcolor}
\usepackage{amsmath}
\usepackage{amssymb}
\usepackage[linesnumbered,ruled,vlined]{algorithm2e}
\usepackage{pseudocode}
\usepackage[english]{babel}
\usepackage{tabularx}
\usepackage{gensymb}
\usepackage{textcomp}
\usepackage{placeins}
\usepackage{balance}
\usepackage{booktabs}

\begin{document}




\title{MyWear: A Smart Wear for Continuous Body Vital Monitoring and Emergency Alert}

\author{
\begin{tabular}{cc}
Sibi C. Sethuraman & Pranav Kompally \\
School of  Computer Science and Engineering,  &  School of  Computer Science and Engineering \\
VIT-AP University, AP 522237, India. & VIT-AP University, AP 522237, India.\\
Email: chakkaravarthy.sibi@vitap.ac.in & Email: pkompally@gmail.com
\\\\
Saraju P. Mohanty &  Uma Choppali  \\
Dept. of Computer Science and Engineering & School of ETMS \\
University of North Texas, USA. & Dallas College, Mesquite, TX 75150, USA.\\
Email: saraju.mohanty@unt.edu & Email: uchoppali@dcccd.edu
\\
\end{tabular}
}

\maketitle

\begin{abstract}
Smart healthcare which is built as healthcare Cyber-Physical System (H-CPS) from Internet-of-Medical-Things (IoMT) is becoming more important than before. Medical devices and their connectivity through Internet with alongwith the electronics health record (EHR) and AI analytics making H-CPS possible. IoMT-end devices like wearables and implantables are key for H-CPS based smart healthcare. Smart garment is a specific wearable which can be used for smart healthcare. There are various smart garments that help users to monitor their body vitals in real-time. Many commercially available garments collect the vital data and transmit it to the mobile application for visualization. However, these don't perform real-time analysis for the user to comprehend their health conditions. Also, such garments are not included with an alert system to alert users and contacts in case of emergency. In MyWear, we propose a wearable body vital monitoring garment that captures physiological data and automatically analyses such heart rate, stress level, muscle activity to detect abnormalities. A copy of the physiological data is transmitted to the cloud for detecting any abnormalities in heart beats and predict any potential heart failure in future. We also propose a deep neural network (DNN) model that automatically classifies abnormal heart beat and potential heart failure. For immediate assistance in such a situation, we propose an alert system that sends an alert message to nearby medical officials. The proposed MyWear has an average accuracy of 96.9\% and precision of 97.3\% for detection of the abnormalities.
\end{abstract}

\begin{IEEEkeywords}
Smart Healthcare, Healthcare Cyber-Physical System (H-CPS), Internet-of-Medical-Things (IoMT), Smart Garment, Smart Home, Heart Condition, Stress, Medical Data Security.
\end{IEEEkeywords}

\section{Introduction}
\label{Sec:Introduction}

The Internet-of-Medical-Things (IoMT) based healthcare Cyber-Physical System (H-CPS) has made smart healthcare possible with enhanced quality of care and faster diagnosis \cite{Aazam_MCE_2020-Mar, JBHI.2020.2973467, Ghamari_Sensors_2016-Jun}.
Smart healthcare is now omnipresent, in the form wearables and implantabale devices with connectivity to have perpetual and continuous body vital monitoring, as it has significant capability to improve quality human life \cite{Joshi_MCE.2020.3018775, Hsu_MCE_2020-Jan, MCE.2020.2969202}. It is well understood that IoMT based H-CPS that makes smart healthcare is more important with the current pandemic scenario. With the introduction of the IoMT, integration of multiple sensors has enabled applications and devices to work efficiently and effectively. IoMT devices are usually connected to the network either by wired or wireless communication means. Smart healthcare is already being effective in fitness tracking and has potentials to revolutionize many aspects, from continuous body health monitoring to performing automatic analysis of electrocardiogram (ECG/EKG) and computed tomography (CT) scan. Smart healthcare can allow doctors to remotely monitor patients based on their fitness tracker data \cite{thomson_nuss_2019, MCE.2019.2956205, Zhu_MCE_2019-Sep}. Wearable such as smartwatches allow users to record ECG/EKG from the wrist and share a copy of the report to doctors for assistance and opinion. Most of the hospitals around the globe use body vital systems that backup the data in real-time in the cloud for doctors to monitor by sitting at home. IoMT-based H-CPS helps in collecting several vitals that describe the health of the person and transmit them to the cloud for post processing. The introduction of IoMT in wearable devices has unlocked various applications in areas such as self healthcare, fitness and yoga.

In the world of wearable technology, fitness trackers have been the part of our lives. Millions of people around the globe use wrist-worn fitness trackers \cite{Gonzales_2019}. People are using fitness trackers to keep track of their daily activities and fitness routine. Millions use wearable IoMT technology that keeps track of both physical and mental health \cite{Joshi_TCE.2020.3011966, Jain_MCE_2020-Jan, 1932296818768618}. In many cases the use of home based healthcare monitoring devices require the user to be rested and the devices used to record vitals. With the upcoming technologies such as bio-medical textiles and devices that can be worn by the user, making him/her free to move around. Many such wearable garments are being used by sports teams and athletes to improve their performance by analyzing body musculoskeletal data recorded by the garment \cite{Pandian_2008}.

\begin{figure}[htbp]
	\centering
	\includegraphics[width=0.65\textwidth]{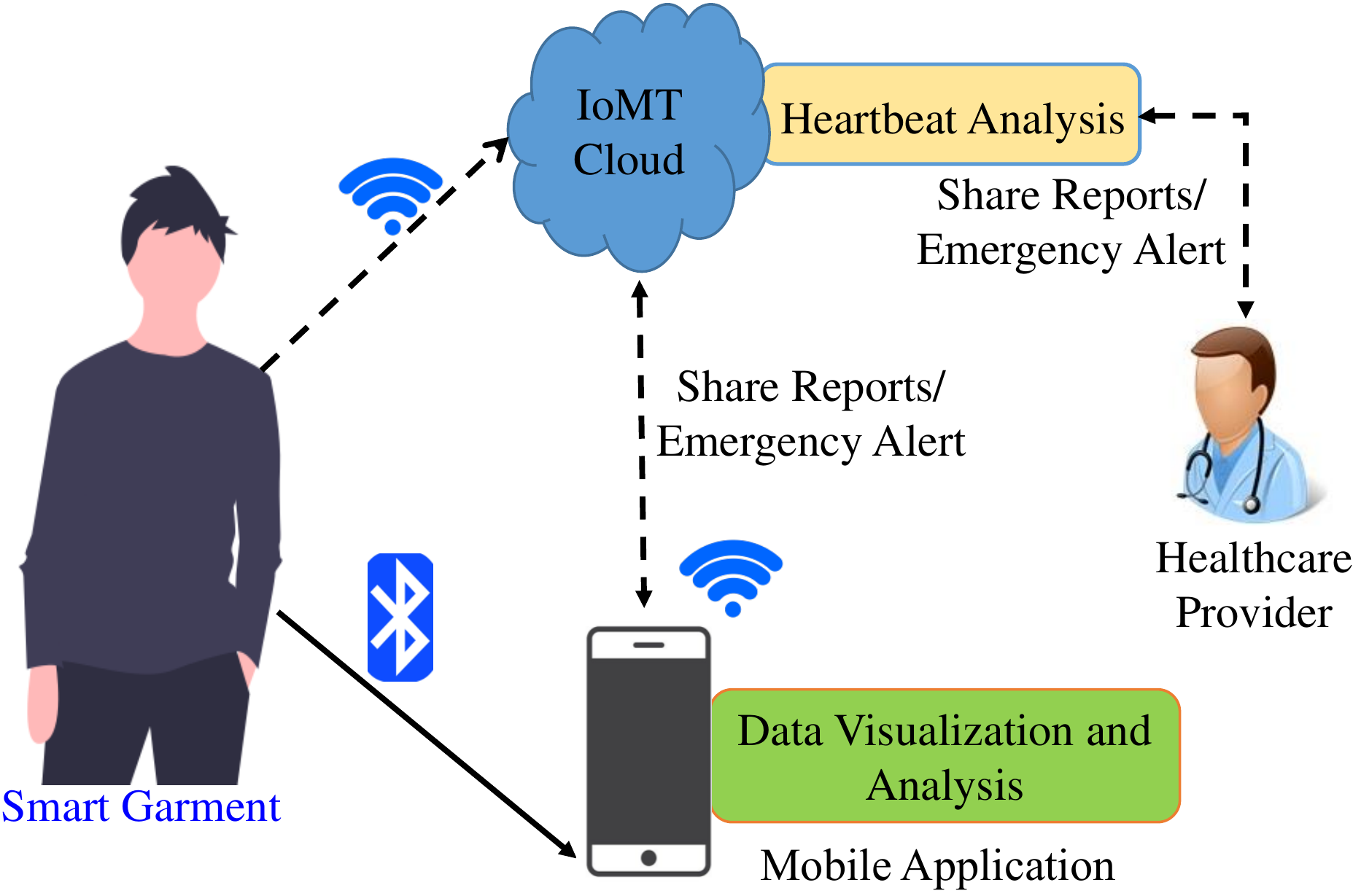}
	\caption{A thematic overview of the proposed MyWear.}
	\label{FIG:Thematic_Concept_Architecture}
\end{figure}

Many fitness enthusiasts are using garments that have embedded sensors to suggest whether the workouts or yoga is performed efficiently. These garments also analyze physiological data with equipped GPS modules that track the user's location \cite{JSEN.2019.2949608, JERM.2019.2929676}. With body vital data, fitness trainers, yoga instructors and doctors can modify the workout plan to favor a particular individual. The same data can be used for nutritionists and dietitians to analyze the body metabolism and accordingly plan the diet. Wearable garments are being developed not only to monitor ECG but also the activity the user is performing while wearing them \cite{Lee_Young_2009}. Law enforcement personnel and armed forces are using wearable medical devices to analyze the stress and health of the soldiers and policemen throughout the day. The data collected provides insights on the lifestyle of the person. In the current work, we envision an wearable called MyWear which is depicted in Fig. \ref{FIG:Thematic_Concept_Architecture} that can continuously monitor stress, heart conditions, muscle activities, and fall. With detailed analysis of the body and its muscle activity, analysts are helping athletes and teams perform and build better.

Rest of the paper is organized as follows: Section \ref{Sec:Contributions} explains the contributions made in this paper. Section \ref{Sec:Related_Research} discusses about the existing related research. Section \ref{Sec:System_architecture} provides the system level architecture. Section \ref{Sec:Stress_Monitoring_Method} discusses the method proposed for automatic heart rate and stress. Section \ref{Sec:Heart_Arrhythmia_Detection_Method}  presents our method for detecting heart arrhythmia. Section \ref{Sec:Fall_Detection_Method}  presents our method for automatic detection of fall. Section \ref{Sec:Experimental_Validation} explains the specific design of the proposed MyWear system along with the its validation through the results of the proposed MyWear system and also provides a comparative analysis with the state of the art similar wearables. Finally, the paper concludes in Section \ref{Sec:Conclusion} with brief discussions on future directions.

\section{Related Prior Research}
\label{Sec:Related_Research}

Consumer electronics to build smart healthcare is an active research area as evident from the fact that we see increasing more healthcare features are available in wearable and smart phones. The research on consumer electronics for smart healthcare has been undertaken in many fronts  including stress management \cite{Rachakonda_TCE_2019}, diet management \cite{Rachakonda_TCE_2020-May}, assisting visually impaired individuals \cite{MCE.2018.2797741} and wearables focusing on women's health \cite{Ava_smart_bracelet}, hearing aids \cite{Lin_Lai_2018}, and garments \cite{Foroughi_2016}. A consumer electronic device that can automatically quantify calorie intake as well as stress of an user is available \cite{Rachakonda_TCE_2020-May}. Heart rate estimation using a photoplethysmography (PPG) based device was presented in \cite{Puranik_TCE_2020-Jan} that deployed neural networks. A framework that can automatically monitor stress level from the physical activities was proposed in \cite{Rachakonda_TCE_2019}. A smart watch for continuous monitoring of data in a privacy-assured manner was presented in \cite{Kim_TCE_2019-Aug}. A ECG signal analysis method using discrete cosine transformation (DCT) has been presented in \cite{Raj_Ray_2018}. A semi-automated IoMT-enabled diet management was discussed in \cite{Prabha_Saraju_2018}. A framework for detection of elderly fall and ECG abormality detection was presented in \cite{Wang_TCE_2016-May}.

\begin{table*}[htbp]
	\centering
	\caption{MyWear as compared to similar works in consumer electronics.}
	\label{TBL:Product_Comparison}
\begin{tabular}{|p{1.8cm}|p{1.2cm}|p{1.5cm}|p{2.0cm}|p{1.4cm}|p{1.2cm}|p{1.2cm}|p{1.2cm}|p{1.0cm}|}
		\hline 
		Consumer Electronics		& Real-Time HRV & Muscle Activity Detection & Abnormal Heartbeat Detection	& Stress Level Detection & Fall Detection & Fall Prediction & Built-in Alert & Data Security
		\\
		\hline
		\hline
Puranik et al. \cite{Puranik_TCE_2020-Jan} &  Yes &  No	& No	& No 	& No	& No	&No	& No\\ 
\hline
		Garment in \cite{Hexoskin_2020}	&Yes	&No	&No	&  Yes & Yes	& No &No	&No   \\
	\hline
Raj et al. \cite{Raj_Ray_2018} & Yes  & No	& Yes	& No	& No	& No & No	&	No \\
\hline
		Garment  in \cite{Athos_2018}	&No	&Yes	&No		&Yes & Yes & No	&No	&No \\
\hline
Wang et al. \cite{Wang_TCE_2016-May} & Yes  & 	No & Yes	& No	& Yes & No	& No	&	No  \\
\hline
Farjadian et al. \cite{Farjadian_ICORR_2013} &  No & Yes	& No	& No	& No & No	& No & No	\\
\hline
Pandian et al. \cite{Pandian_2008} & No  & No	& No	& No	& No	& No & No & No	\\
\hline
		\textbf{MyWear (Current Paper)}	&Yes	&Yes	&Yes 	&Yes	&Yes &Yes	&Yes &Yes\\
		\hline
	\end{tabular}
\end{table*}%

Table \ref{TBL:Product_Comparison} presents a comparative perspective of similar consumer electronics as the proposed MyWear of the current work.  
There are few smart wearable garments that can monitoring human body vitals in real-time \cite{Foroughi_2016}. Smart wearable garment in \cite{Athos_2018} uses surface electromyography (sEMG) to analyze the intensity of muscle activity of athletes.  
However, there is no HRV analysis observed in the training system to detect stress levels of the user. A consumer products \cite{Hexoskin_2020}  analyzes sleep activity and ECG for heart rate variability (HRV) analysis. 
However, it does not examine the muscle activity of the user. 
Garment in \cite{Farjadian_ICORR_2013} use electromyography (EMG)  to detect muscle activity and assist the user in physical therapy.  
There are few proposed solutions that use accelerometer data to analyze the intensity of exercises, however, it cannot be used in calculating individual muscle activity in comparison to using sEMG. Moreover, no accurate measurement of the body orientation of the user is observed in the above mentioned wearables. There was no alert system observed in the above-mentioned garments, that notifies users contacts, paramedic forces in case of emergency after detecting any abnormalities.

\section{Contributions of Current Paper}
\label{Sec:Contributions}

To the knowledge of the authors {MyWear is the first smart garment to introduce an integrated mechanism for automatic} HRV analysis, stress analysis, muscular activity analysis, and alert system to seek assistance in emergencies while providing data security.

\subsection{Problem Formulation}

We intend to address the following health conditions through the research and development of MyWear:
\begin{itemize}	
\item
How can health condition be automatically monitored and healthcare provider be notified?

\item 
How can heart abnormalities be automatically detected from the ECG signal?

\item 
How can stress be automatically detected from the ECG signal?

\item 
How can muscle activity be automatically detected?

	\item 
How can fall of an individual be automatically detected?

\end{itemize}

\subsection{Proposed Solution of the Current Paper}

MyWear provides following novel solutions to the research objective stated above:
\begin{itemize}
	\item 
An automated continuous health monitoring system by collecting body vitals in regular intervals of time, store it in a secured fashion and alert healthcare providers.

	\item 
A cloud based platform for user and healthcare providers to monitor previous logs and current routine in real-time.
	
		\item 
A novel automatic method to automatically analyze stress levels using Electrocardiogram (ECG) signal in real-time.
	
	\item 
A novel method to detect heart abnormality arrhythmia using Electrocardiogram (ECG) signal in real-time.
	
\item 
A novel automatic method to determine fall of the user.
\end{itemize}

\subsection{Novelty and Significance of the Proposed Solution}

The novelty and significance of the proposed solutions include the following:
\begin{itemize}
\item
A novel wearable that can automatically perform heart rate monitoring, heart arrhythmia, stress monitoring, and fall detection for a complete healthcare solution. 

\item 
A secured smart garment with an automated body vital data logging system to the cloud and mobile application.

\item 
ECG signal analysis using deep learning model to detect different types of abnormalities in heartbeat that is not observed in any of the smart garments. 

\item 
Real-time stress detection using Heart Rate Variability (HRV) from ECG which is not present in most of the smart garments.

\end{itemize}

\section{System Level Architecture of the Proposed Framework}
\label{Sec:System_architecture}

We present our vision of Internet-of-Medical-Things (IoMT) based healthcare Cyber-Physical System (H-CPS) with the proposed smart garment MyWear integrated in it. MyWear can be with 3 different options depending on where the computation happens and where the intelligence is built in \cite{Rachakonda_TCE_2020-May, MCE.2017.2776462, MCE.2017.2714695}: (1) IoT-end device in-sensor computing, (2) IoMT-edge/fog computing, and (3) IoMT-cloud computing paradigm. In option-1, IoMT-end device with all the sensors and models integrated in it, while IoMT-cloud stores data and diagnostic results \cite{8719325}. In this option, IoMT-end device need to have a bit higher level of computational capability and battery life than the usual sensors to present results instantaneously to the user. Also, with the use of light-duty machine learning models (i.e. TinyML) it can be possible with limited computational and battery resources at the IoMT-end device \cite{Mohanty_VAIBHAV_2020_Panel}. In Option-2, the ML models are part of IoMT-edge devices like edge routers and edge datacenters (EDC) for faster response to the users \cite{8684800}. In Option-3, heavy-duty accurate ML models can run in IoMT-cloud to detect the health conditions. Thus, Option-2 with IoMT-edge is a good trade-off accuracy and fast response.  A more effective option is to have TinyML models in IoMT-sensor for fast response and at the same time have another health condition evaluation at IoMT-cloud before sending alarm to healthcare providers.

\subsection{Proposed Healthcare Cyber-Physical System (H-CPS) using our Smart Garment}

\begin{figure}[htbp]
\centering
\includegraphics[width=0.85\textwidth]{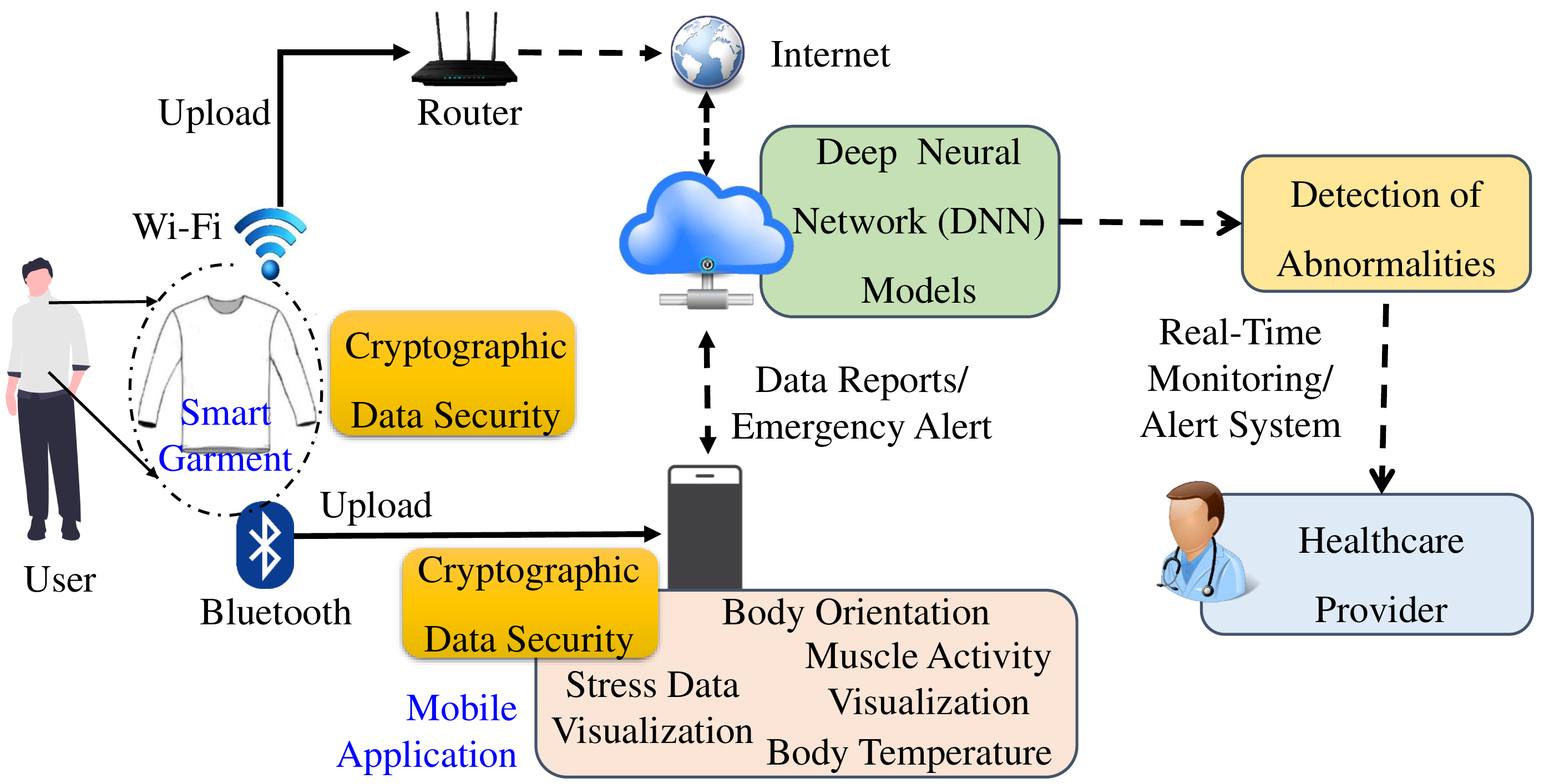}
\caption{A detailed architecture of the proposed MyWear in a Healthcare Cyber-Physical System (H-CPS) framework.}
\label{FIG:Smart-Garment_Framework_in_H-CPS}
\end{figure}

The complete overview of the proposed system in H-CPS framework is depicted in Fig. \ref{FIG:Smart-Garment_Framework_in_H-CPS}. The garment acts as the IoMT-End device and the input point for the mobile application and cloud service. Surface dry electrodes are embedded in the garment keeping in mind the optimal position to obtain stable signals. The dry electrodes are connected to the respective ECG and EMG sensors. These sensors extract, amplify and filter the raw signals therefore removing noise and unwanted artifacts. The filtered data is sampled by sampling unit along with the data received from the temperature and Inertial Measurement Unit (IMU) sensors. The temperature sensor measures the body temperature whereas IMU sensors measure the change in body orientation. Collectively, data is transmitted to the mobile application and cloud for further analysis using the embedded Bluetooth and Wi-Fi module, respectively. The vital data is AES128 encrypted and can only be decrypted or accessed in the user's mobile application. Thus, keep the body vital data safe and secure, only readable by the owner/user. The mobile device displays ECG in real-time along with the stress level of the user. The mobile application visualizes muscle activity in different muscle regions on the human map pertaining to the individual along with the body orientation and body temperature. The application allows the user to share Data with medical officials for further analysis and assistance. Meanwhile, the ECG data transmitted to the cloud is analysed to detect any abnormal Heartbeats. The proposed Deep Learning model deployed in the cloud checks for any abnormalities and detects the kind of abnormality that occurred. In case of emergency, an alert is sent to medical officials for immediate assistance. Moreover, the body vital data received in the cloud can be monitored by medical officials in real-time.

MyWear gathers body vitals such as Heart Rate, body temperature, muscle activity and sends it to IoMT-edge and IoMT-cloud through Internet. The smartphone acts as an interface to visualize data post analysis for user’s information. The deep learning (DL) model in the cloud analyzes the user's ECG and stress level. It also returns a report to the user's smartphone for the user to access. 
The \textit{main objectives of MyWear} are the following:
\begin{itemize}
\item	
To create an automated health monitoring wearable that analyzes the user's body vitals regularly.
\item	
To provide a solution that analyzes a user's stress level based on Electrocardiogram.
\item	
To bridge the communication between User and medical officials, real-time user monitoring system to allow doctors, therapists to analyze the user's routines.
\item	
To create an alert system to call for help in case of emergency.
\end{itemize}

\subsection{Architecture of the Proposed Smart Garment}

\begin{figure}[htbp]
	\centering
	\includegraphics[width=0.75\textwidth]{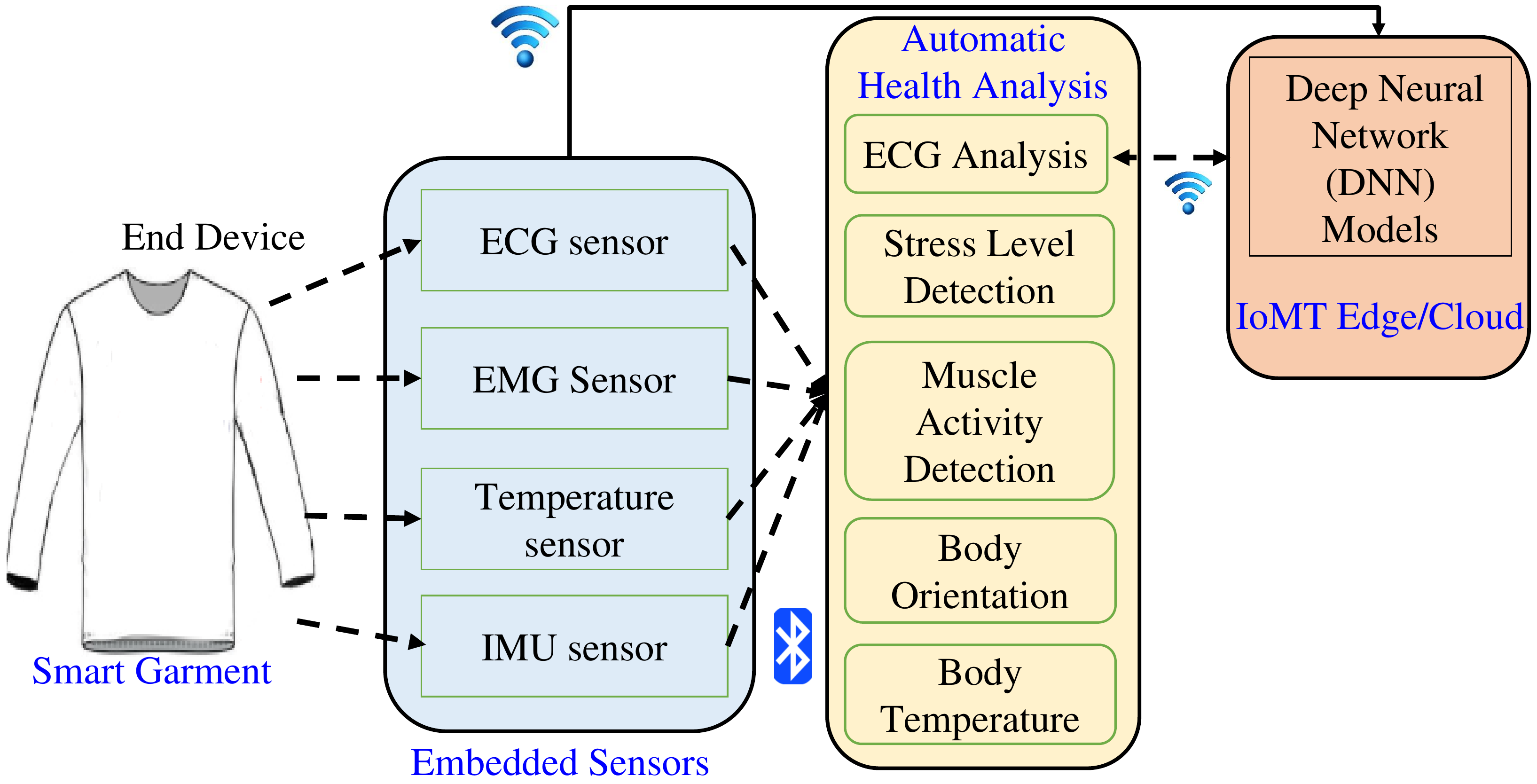}
	\caption{Architecture of the proposed Smart Garment (MyWear).}
	\label{FIG:Proposed_Smart_Garment_Architecture}
\end{figure}

Fig. \ref{FIG:Proposed_Smart_Garment_Architecture} shows the architecture of the proposed smart garment. The garment (called MyWear) is embedded with dry electrodes that are connected to ECG and EMG sensors. The sensors extract, amplify and filter the raw signals captured by the dry electrodes. The filtered data is sampled by sampling unit along with the data received from Temperature and IMU sensors. The data is then transmitted to the smartphone and cloud server for storage and analysis. The received ECG data is used to analyze user stress levels and heart rate as discussed later. The analyzed outcome along with a report is sent to the smartphone application for the user to comprehend. The filtered muscle data received at the smartphone application end is used to visualize on the human map. The application also displays the body orientation and temperature of the user. The user can also grant access to cloud data for doctors, therapists and anyone to monitor on a real-time basis. In case of emergency, an alert message is sent to contacts and paramedic forces for immediate assistance.

\subsection{Electrocardiogram (ECG) Acquisition Unit for Heart Rate}

\begin{figure} [htbp]
\centering
\subfigure[Placement of electrodes and Einthoven's Triangle.]{
	\centering
	\includegraphics[width=0.35\textwidth]{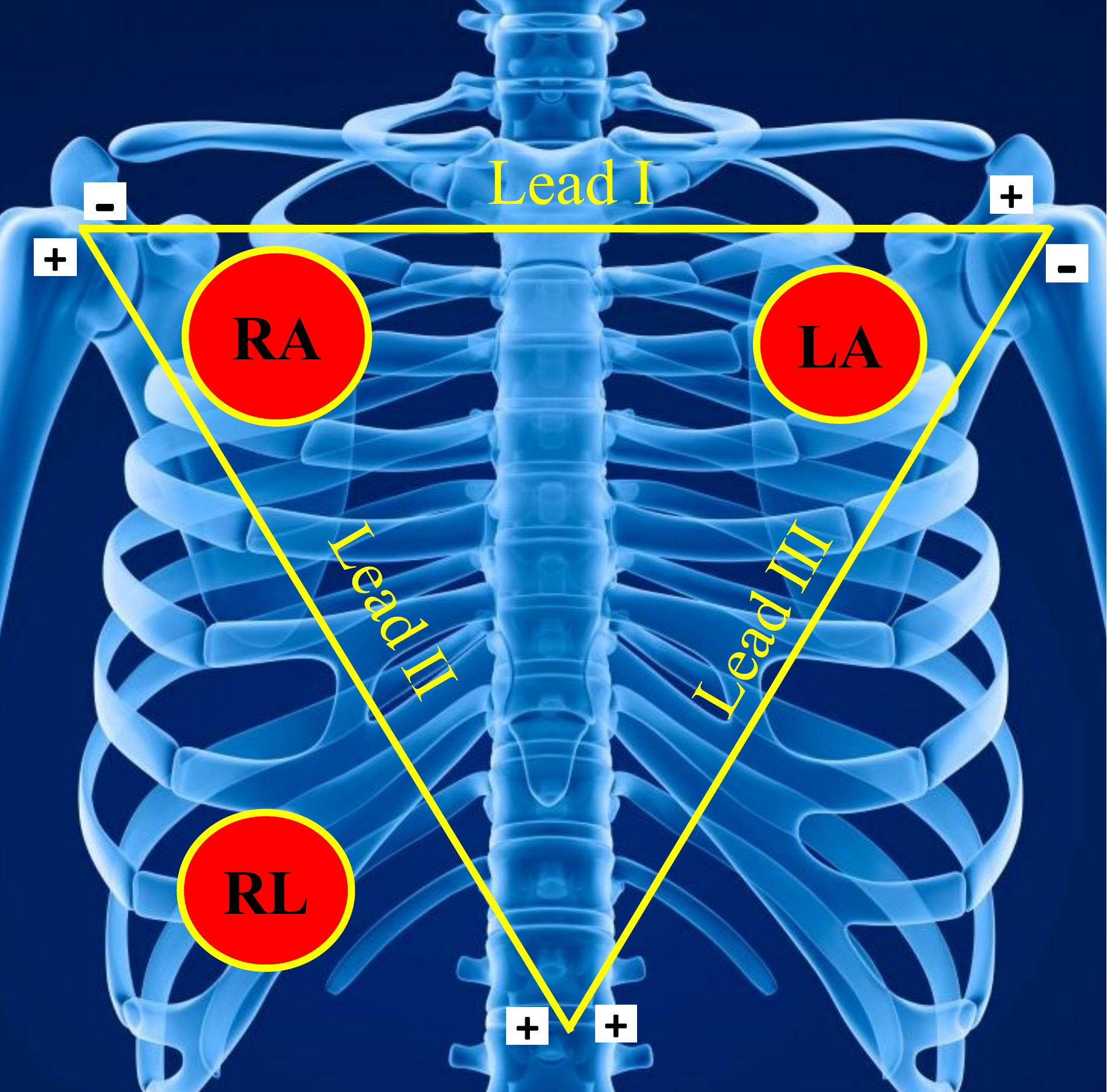}
        	\label{FIG:Einthoven_Triangle} 
}
\subfigure[P segment, QRS complex and T segment of Electrocardiogram.]{
	\includegraphics[width=0.55\textwidth]{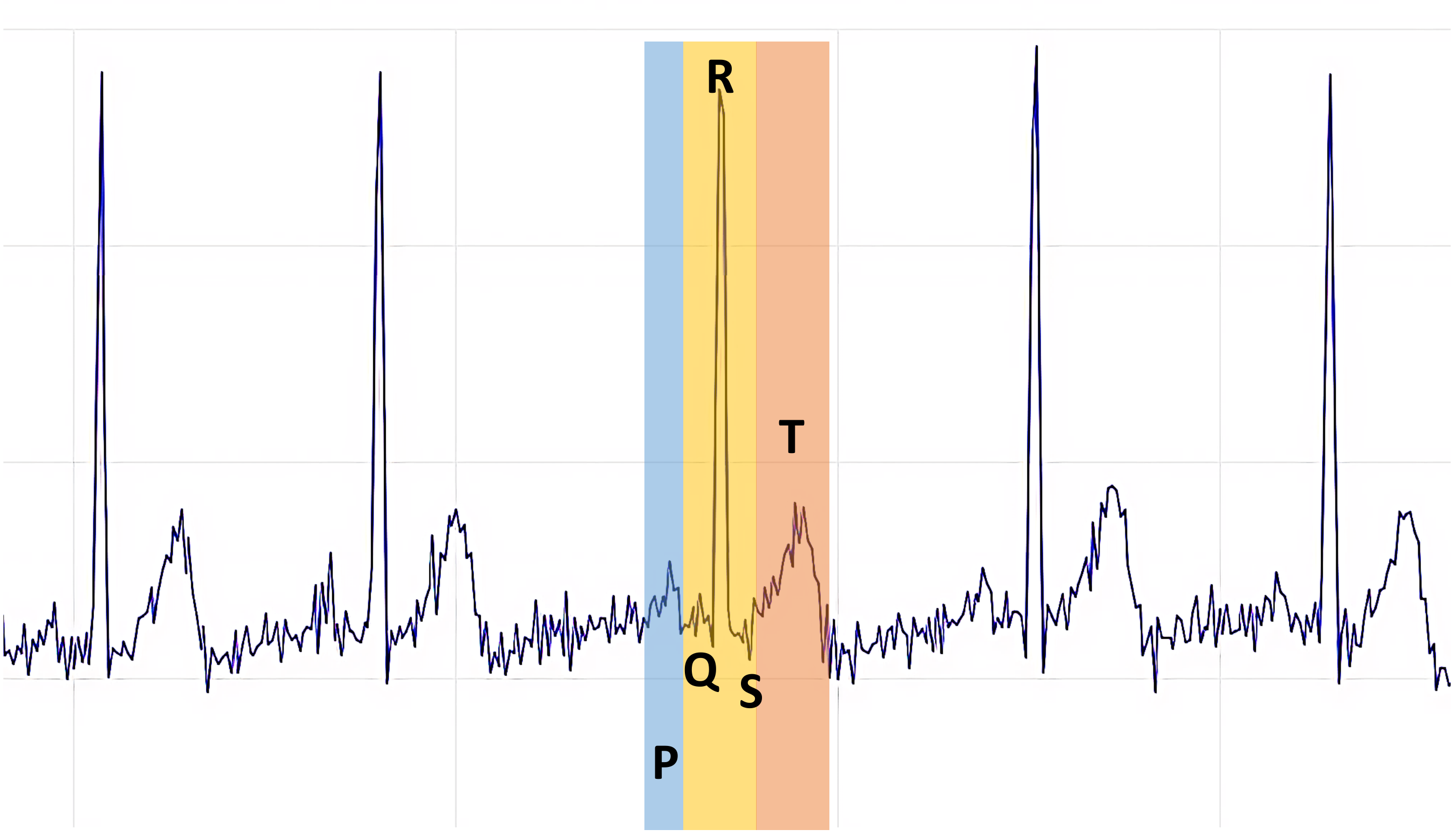}
	\label{FIG:ECG_Segment_PQRST}
}
\caption{Einthoven's Triangle based electrode placement and P segment, QRS complex and T- segment of ECG.}
\end{figure}

Electrocardiogram (ECG) is a technique used to measure the electrical activity produced by the heart during a diastole and systole (relaxation and contraction). ECG is a reliable method to measure the heart rate variability and beats per minute (BPM) \cite{Iskandar_AISP_2019}. For convenient use, a 3 electrode system is used. The electrodes are placed following the Einthoven's triangle (refer Fig. \ref{FIG:Einthoven_Triangle}), which depicts the placement of electrodes to obtain stable ECG \cite{Kanani_2018, Villegas_2019}.

\subsection{Euler’s Angle Unit Using Inertial Measurement Unit (IMU) Sensor}

MyWear is embedded with an Inertial Measurement Unit (IMU) sensor. It measures and provides details on orientation and velocity of a particular entity with the use of accelerometers and gyroscopes. IMU detects rotational movements and acceleration of the Euler angles, also known as yaw, pitch and roll. The gyroscope measures angular velocity which defines the rate of change of angle with respect to $x$, $y$ and $z$ axes. The accelerometer defines the velocity at which the entity is rotating in a particular axis.

\subsection{Electromyography (EMG) Unit}

Electromyography (EMG) is used to measure the change in electric potential that depicts the force exerted by the muscle. For ease of use, a three-electrode system is used to measure the muscle activity at a particular muscle region. Two electrodes are used to measure the muscle signal and the third electrode acts as a ground. Initially, the sensor is tuned by changing the gain. The gain helps in adjusting the sensitivity of signal acquisition to capture stable signals with low noise. EMG helps in understanding the muscle activity and its intensity in a muscle region. EMG is used to measure the change in electric potential generated at neuromuscular junctions as electric signals or action potential pass through. Clinical settings of EMG use a needle that is inserted into the muscle. For measuring ECG on the go and better ease of use, surface EMG is chosen \cite{Samarawickrama_2018, Alimam_2017}. Surface EMG uses electrodes that measure overall activity of a large portion of the muscle. The activity is measured in voltage and represents the amount of force exerted by the muscle in real-time. MyWear records muscle activity at the biceps or biceps brachii and at the chest or pectoralis major.

\subsection{Body Orientation Detection Unit}

Upon Calibration of the Euler's angle to the initial position of the user, the change in angles is recorded and defined for a few positions such as bending is four directions: right, left and forward, back. The user is prompted with the text in the application that shows the orientation of the user as shown in Fig. \ref{FIG:Mobile_App_for_User}. Body orientation provides insights into a person's activity.

\subsection{Emergency Alert Unit}

The ECG data is sent to the model to detect any abnormalities. The proposed deep Learning model detects any abnormalities if any. Detected abnormalities are sent as an alert to the user’s mobile application and the medical official. Regular intervals of Abnormalities are usually considered to cause a potential heart failure and hence, a prompt is sent to the user’s smartphone application and triggers an alarm. Another alert is sent to Medical officials and doctors for immediate assistance.

\subsection{Data Security Unit}

Privacy and Security play a major role when handling Body vital or Health Data. Health data when misused can pose serious privacy and security violations and eavesdropping. In order to keep user's data safe and secure, the body vitals data is AES128 encryption enabled. Advanced Encryption Standards (AES) is an encryption method which uses a 128-bit block with key size of 128bits. It functions on a symmetric algorithm wherein the key is used to encrypt and decrypt data. Before transmitting data from Arduino to Application. The collected data from all sensors is encrypted using AES128 with the predefined key. The key is unique to a particular garment, mobile device and is assigned by the mobile application when pairing and registering initially. Any other device with installed application will not be able to access and receive the sensors data. Once the encrypted data is transferred to assigned application, the data is decrypted with the pre-defined key.

\section{Proposed Methods for Automatic Heart Rate and Stress Monitoring from ECG}
\label{Sec:Stress_Monitoring_Method}

\subsection{Proposed Method to Obtain Heart Rate from ECG}

Fig. \ref{FIG:ECG_Segment_PQRST} shows what an ECG graph would look like and its features. Every beat of the heart corresponds to a P-QRS-T waveform in the graph and collection of multiple waveforms depicts the successive beats in a period of time. As seen in Fig. \ref{FIG:ECG_Segment_PQRST}, the `P' wave or bump that precedes the QRS complex represents the depolarization of the atria which tends to brief isoelectric period or state of near zero voltage.

The P wave lasts no more than 0.10 seconds and 2.4 mm tall. P wave is followed by the QRS complex. QRS complex consists the rapid succession of three waves i.e. `Q' wave, `R' wave and `S' wave. The QRS complex represents the activation of ventricular muscles that contract the heart. The normal duration of the QRS complex is between 90 milliseconds and 100 milliseconds. The R wave is usually a peak and positive unlike Q and R wave. The `T' wave follows the QRS complex and indicates ventricular repolarization that leads to relaxation of the heart. This repeats for every single beat of the heart. The heart rate is measured in beats per minute as the following:
\begin{equation}
\text{Heart Rate } (bpm) = \left( \frac{60}{T_r} \right),
\end{equation}
where $T_r$  = time between two successive $R$ peaks.

To calculate the time between two successive `R' peaks, the time at which first and second peaks occurred is saved. Subtracting the first peak time from second peak time results in RR-interval time. Beats per minute are obtained as follows:
\begin{equation}
\text{Heart Rate } (bpm) = \left( \frac{1.0}{RRinterval} \right) \times 60.0 \times 1000,
\end{equation}
where $RR$ intervals are used for Heart Rate Variability (HRV) analysis to detect the stress levels of the user.

\subsection{Metrics for obtaining Heart Rate Variability (HRV) score}

\subsubsection{Time-domain metrics}
HRV score can be calculated by measuring the time between two successive RR intervals using Time domain metrics as shown below:

\paragraph{MeanRR}
Average of all $RR$ intervals (distance between two `R' peaks) is calculated using the following expression:
\begin{equation}
MeanRR = \left(  \frac{\sum\limits ^{n}_{i=1} R_{i}}{n} \right).
\label{eq:1}
\end{equation}

\paragraph{Standard Deviation of RR intervals (SDNN)}

The standard deviation of RR intervals (also known as NN-interval) is calculated by using the following expression:
\begin{equation}
SDNN\ =\ \sqrt{\frac{\sum\limits ^{n}_{i=1}( R_{i} -mean)^{2}}{n}} .
\label{eq:2}
\end{equation}

\paragraph{Root Mean Square of Successive Differences (RMSSD)}
The root mean square of two RR  intervals' differences is calculated using the following expression:
\begin{equation}
RMSSD\ \ =\ \sqrt{\frac{\sum\limits ^{n-1}_{i=1}( R_{i} -R_{i+1})^{2}}{n-1}} .
\label{eq:3}
\end{equation}

\subsubsection{Frequency-domain metrics}
HRV score can be calculated by measuring the low and high frequency heart beats. Poincare plot is a reliable frequency domain metric among other metrics to visualize HRV. The algorithm plots two successive RR intervals and depicts how well each RR interval predicts the consecutive RR interval.

\subsection{Proposed Method to Automatically Monitor Stress from ECG}
\begin{figure}[htbp]
	\centering
	\includegraphics[width=0.80\textwidth]{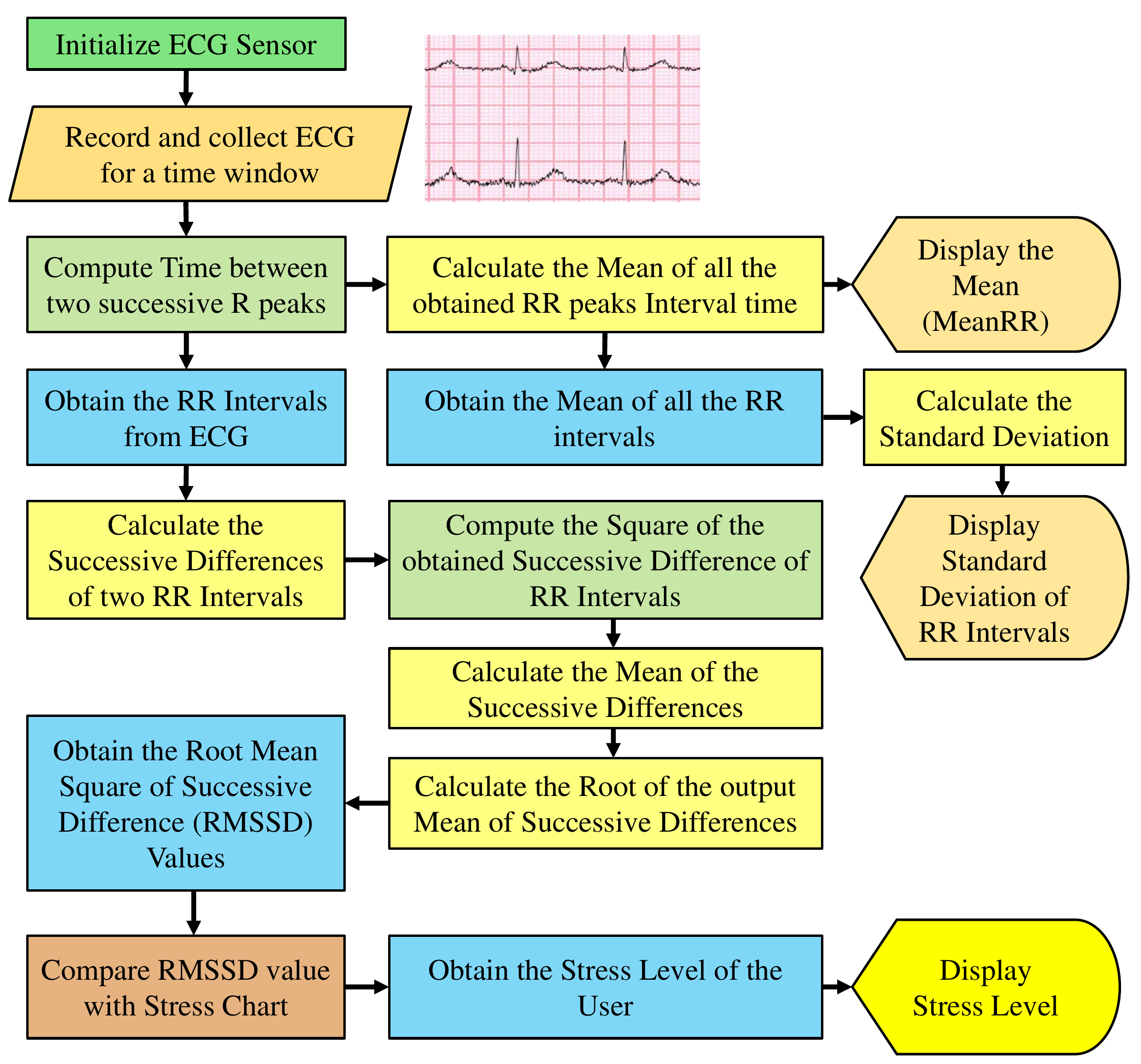}
	\caption{Proposed approach for calculating stress level from the heart rate.}
	\label{FIG:Proposed_Stress_Level_Calculation_Flow}
\end{figure}

Fig. \ref{FIG:Proposed_Stress_Level_Calculation_Flow} shows the algorithm to calculate stress level using Heart rate. RMSSD is usually obtained from ECG and is considered as the HRV score  \cite{Wu_IS3C_2016}. Studies have shown that an increase in HRV depicts a reduction in stress levels \cite{Wang_mse_2012} and vice versa. High HRV score was found in users performing optimal levels of fitness routine. However, sometimes, during strenuous exercise, low HRV scores were noted because of the body trying to increase the user's heart rate for the activity. A user achieves a high HRV score during sleep as the result of state of relaxation and low stress levels. HRV score changes depending on the user's activity. Apart from HRV, ECG can also be used to detect Heart Arrhythmia. Table \ref{TBL:HRV_vs_Stress} shows the relation between HRV score and stress levels  \cite{mindful_HRV_2020}.

\begin{table}[htbp]
\centering
\caption{Relation between HRV score and stress levels \cite{mindful_HRV_2020}.}
\label{TBL:HRV_vs_Stress}
\begin{tabular}{|p{3cm}|p{3cm}|}
 \hline 
\textbf{HRV Score} & \textbf{Stress Level} \\
 \hline
 \hline
90+   & Very low   \\
\hline
80-90  &Low \\
\hline
71-80	&Moderate \\
\hline
61-70   &Average\\
\hline
$<$60 &High \\
\hline
\end{tabular}
\end{table}

\section{Proposed Deep Neural Network (DNN) Model For Automatic Detection of Heart Arrhythmia}
\label{Sec:Heart_Arrhythmia_Detection_Method}

An arrhythmia is a disorder of the heart that affects the rate or rhythm at which the heart beats. This takes place when electrical impulses that direct and regulate heartbeats don't function properly. As a result the heart may be too fast (tachycardia), too slow (bradycardia), too early (premature contraction), or too erratically (fibrillation). Thus, it is of critical importance to determine such conditions automatically and in real-time. We present a Deep Neural Network (DNN) model based method that is integrated in MyWear for the automatic determination of arrhythmia and correspending automatic warning to healthcare provider.

\subsection{Proposed DNN Model for Heart Arrhythmia}

\begin{figure*}[htbp]
	\centering
	\includegraphics[width=0.99\textwidth]{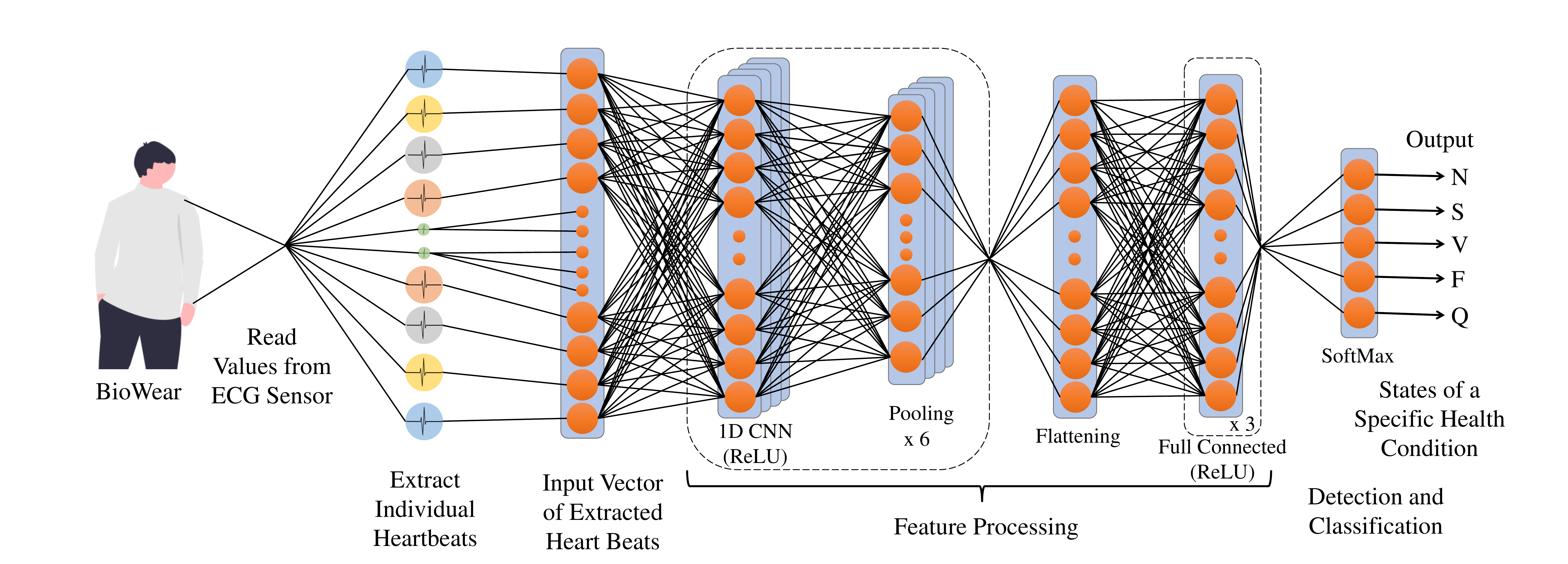}
	\caption{Deep Neural Network (DNN) Model model explored for our MyWear.}
	\label{FIG:DNN_Model_Architecture}
\end{figure*}

Fig. \ref{FIG:DNN_Model_Architecture} shows the architecture of the proposed deep learning model. The proposed learning model consists of 6 one-dimensional convolutional layers with 64 filters each and input stride length of 2 and uses Rectified Linear Unit function (ReLU) activation function. Every convolutional layer is succeeded by the Maxpool layer of pool size 2 and stride size 2.  The set of Convolutional layers is connected to 3 Fully Connected Layers. Finally, a softmax function is used in the output layer to predict probabilities of individual classes. The model is used to classify the dataset into four categories based on the heart beat rhythm and predict whether the input heart beat is normal or consisting of abnormalities. The activation function used in every layer is ReLU which is represented as follows:
\begin{equation}
f(x)  = \left\{
    \begin{array}{l}
      1  \:\:    x\textgreater1\\
      x \:\: x=1 \text{ and} \: \: 0 \\
     0 \: \: x \textless 0
    \end{array}
  \right.
\end{equation}

The output layer connected to the fully connected layer of $n$ neuron. The predicted classification of heartbeats for a sample $x$ is denoted by the Softmax function \cite{Xiangrui_2020} defined as the following expression:
\begin{equation}
f_{n} \ ( x) = \left( \frac{e^{( W_{n} h_{x} \ +\ b_{k})}}{\sum\limits_{j =1} ^{K} e^{( W_{j} h_{x} \ +\ b_{j})}} \right) ( n\ =\ 0,...,n-1) , \\
\end{equation}
where $h_x$ is the feature representation of $x$ extracted from the previous convolutional layer, $W_k$ and $b_k$ from nth neuron in the output layer.

\subsection{Metrics for evaluating the DNN Model}

The metrics used for evaluating the proposed DNN model are precision, recall, accuracy and loss \cite{Rachakonda_TCE_2020-May}. In order to evaluate these metrics, there are four basic errors that describe the metrics which are to be defined. The errors are calculated as follows:
\begin{itemize}
\item 
\textit{True Positive (TP)}: Heart beats that belong  to the truth class and predicted as belonging to the same class
\item	
\textit{True Negative (TN)}: Heart beats that do not belong to the truth class and predicted as not belonging to the same class.
\item	
\textit{False Positive (FP)}: Heart beats that do not belong to the truth class and predicted to not belonging to the same class
\item	
\textit{False Negative (FN)}: Heart beats that belong to the truth class and predicted as not belonging to the same class. 
\end{itemize}

These metrics that are used to evaluate the DNN model are the following:
\begin{itemize}
\item 
\textit{Precision}: The ability of the model to identify the possible heart beats from the input:
\begin{equation}
P\ =\left[ \ \frac{TP}{TP+FP} \ *\ 100\%\ \right].
\end{equation}

\item	
\textit{Recall}: The ability of the model to identify all the relevant heartbeats from the predicted possible heartbeats:
\begin{equation}
R\ =\left[ \ \frac{TP}{TP+FN} \ *\ 100\%\ \right].
\end{equation}

\item	
\textit{Accuracy}: The ratio of correct predictions made by the model to the total number of predictions by the model:
\begin{equation}
\alpha =\ \left[ \ \frac{TP\ +\ TN}{TP+TN\ +\ FP\ +FN} \ *\ 100\%\ \right].
\end{equation}
\end{itemize}

\subsection{Preparation of Dataset}

In order to detect abnormalities in ECG, a dataset with classification of heartbeats into normal beats, Ventricular and Supraventricular beats along with Fusion beats and Unknown beats is considered. The MIT-BIH Arrhythmia Dataset \cite{Moody_2001} is used that contains 48 half an hour excerpts of two channel ECG with over 150,000 samples. The proposed model is trained on 100,000 and tested on approximately 22,000 samples. The sample information that is being considered in MyWear is described in Table \ref{TBL:Dataset_MyWear}.

\begin{table}[htbp]
	\centering
	\caption{Different Heartbeat information considered for classification.}
	\label{TBL:Dataset_MyWear}
	\begin{tabular}{|p{3cm}|p{7.5cm}|}
		\hline 
		\textbf{Characteristics} & \textbf{Specifics} \\
		\hline
		\hline
		Number of samples  & 100,000 \\
		\hline
		Sampling frequency & 125Hz\\
		\hline
		Classes & Normal, Supraventricular, Ventricular, Fusion, Unknown\\
		\hline
	\end{tabular}
\end{table}%

\section{Proposed Methods for Automatic Muscle Activity Detection, and Fall Detection and Prediction}
\label{Sec:Fall_Detection_Method}

Electromyography (EMG) helps in understanding the muscle activity and its intensity in a muscle region. EMG is used to measure the change in electric potential generated at neuromuscular junctions as electric signals or action potential pass through. Clinical settings of EMG use a needle that is inserted into the muscle. For measuring ECG on the go and better ease of use, surface EMG is chosen \cite{Alimam_2017, Samarawickrama_2018}. Surface EMG uses electrodes that measure overall activity of a large portion of the muscle. The activity is measured in voltage and represents the amount of force exerted by the muscle in real-time. MyWear records muscle activity at the biceps or biceps brachii and at the chest or pectoralis major.

\subsection{Proposed Method for Muscle Activity Automatic Detection from Electromyography (EMG)}

EMG is used to measure the change in electric potential that depicts the force exerted by the muscle. For ease of use, a three-electrode system is used to measure the muscle activity at a particular muscle region. Initially, to capture stable signals with low noise, the sensor is tuned by changing the gain. The gain helps in adjusting the sensitivity of signal acquisition. EMG helps in understanding the muscle activity and its intensity in a muscle region. EMG is used to measure the change in electric potential generated at neuromuscular junctions as electric signals or action potential pass through. Clinical settings of EMG use a needle that is inserted into the muscle. For measuring ECG on the go and better ease of use, surface EMG is chosen \cite{Alimam_2017, Samarawickrama_2018}. Surface EMG uses electrodes that measure overall activity of a large portion of the muscle. The activity is measured in voltage and represents the amount of force exerted by the muscle in real-time. MyWear records muscle activity at the biceps or biceps brachii and at the chest or pectoralis major.

\subsection{Proposed Method to Calculate Body Orientation}

In order to determine the body orientation. The sensor is calibrated with the user's initial position and orientation. The increase or decrease in the Euler's angle determines the change in body orientation of the user. Upon initializing the sensor, it provides the $X$, $Y$ and $Z$ values, however, these values depend on the sensitivity. The default sensitivity is -2$g$ to +2$g$. To calibrate the sensor, offset values are initialized. Initial position or offset values are recorded when the person is stood up straight and still. These values are written in $X$, $Y$ and $Z$ axis offset registers. After calibration, the user's movements are measured as $X_{out}$, $Y_{out}$ and $Z_{out}$. Roll, Pitch and Yaw values are calculated  using the following expressions \cite{husstech_2020}:
\begin{eqnarray}
Roll, \rho =  arctan \left(\frac{Y_{out}}{\sqrt{( X_{out})^{2} + ( Z_{out})^{2} \ }}\right)
\\
Pitch, \phi  = arctan\ \left(\frac{X_{out}}{\sqrt{( Y_{out})^{2} \ +( Z_{out})^{2} \ }}\right) \\
Yaw, \theta = arctan\ \left(\frac{\sqrt{( Y_{out})^{2} \ +( X_{out})^{2}}}{Z_{out}}\right) 
\end{eqnarray}
These values define the change in the $X$, $Y$ and $Z$ values from the calibrated values. The orientation of the user is provided as simple text showing whether the person is bending towards right, forward or diagonally towards left feet.
The above expressions form the basis of calculating any movement recorded by the IMU as well as computing these would be helpful to detect action such as fall and body orientation.

\subsection{Proposed Method for Fall Prediction}

In order to detect sudden fall of the person triggered by involuntary force, simple three-layer CNN followed by two MaxPooling layers and an output Softmax function is used to predict a probable fall \cite{Zhou_2018}. The model predicts whether the person is about to fall by the change in resultant acceleration obtained from the garment's Accelerometer. The resultant acceleration is calculated using the following expression \cite{Fan_2020}:
\begin{equation}
	g_{_{i}} = \sqrt{\frac{x^{2}_{i} +y^{2}_{i} +\ z^{2}_{i} \ }{g}},
\end{equation}
where $g$ is the acceleration due to gravity which is 9.8. $g_i$ is the resultant acceleration at instance $i, x_i,y_i,z_i$ are the values of accelerations at instance $i$ along $x$, $y$ and $z$ axis, respectively \cite{Fan_2020}. $g_i$ is calculated at every instance of time that accelerometer is received.

\subsection{Proposed Method for Fall Detection}

\begin{figure}[htbp]
	\centering
	\includegraphics[width=0.65\textwidth]{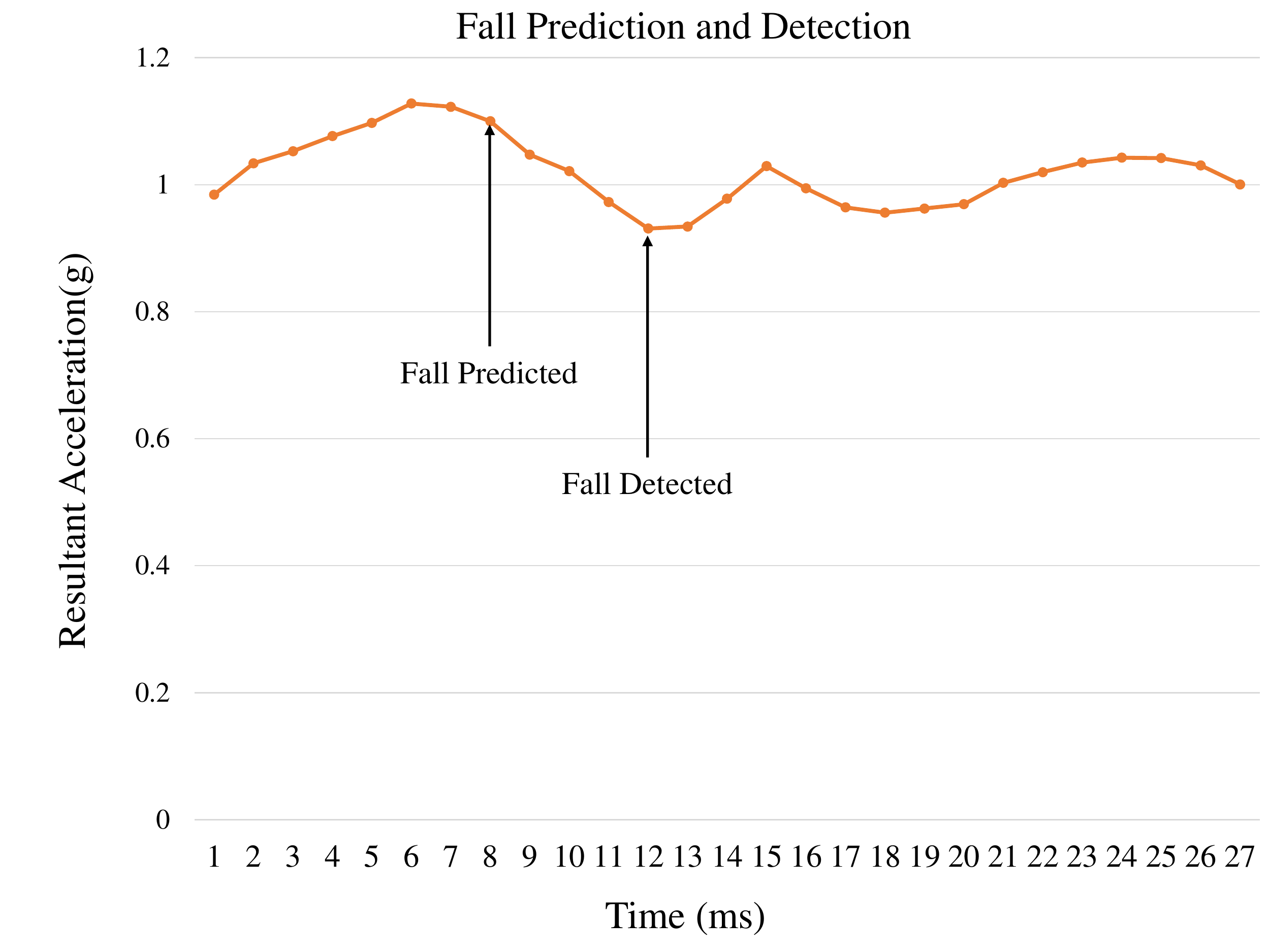}
	\caption{Fall detection and Prediction.}
	\label{FIG:Results_Fall_detection_prediction}
\end{figure}

After a fall is predicted, if it found that the resultant acceleration swiftly decreases below the maximum threshold of $+0.90g$ from $+1g$ in less and quickly increases over $+1g$ in less than 0.3 seconds \cite{Mezghan_2017, HEMALATHA_2013}, it can be concluded that the predicted fall occurred and is detected as seen in the Fig. \ref{FIG:Results_Fall_detection_prediction}. Table \ref{TBL:Body_Vitals_during_Fall} shows the change in body vitals during fall.

\begin{table}[htbp]
	\centering
	\caption{Body Vitals during Fall.}
	\label{TBL:Body_Vitals_during_Fall}
	\begin{tabular}{|p{3cm}|p{7cm}|}
		\hline 
		\textbf{Characteristics} &\textbf{Specifics} \\
		\hline
		\hline
		Fall Prediction  &Swift decrease in Resultant Acceleration(g) below the calibrated threshold \\
		\hline
		Muscle Activity & Quick activation in bicep and Chest Muscle forces\\
		\hline
		Beats/min & A sudden fall triggers quick increase in Heart Rate\\
		\hline
		Fall Detection & Increase in g to above +1g within 0.3 seconds \\
		\hline
	\end{tabular}
\end{table}%

\section{Experimental Validation of MyWear}
\label{Sec:Experimental_Validation}

We prototyped NyWear using off the shelf components. The prototype was validated using various datasets. This Section discusses all the details.

\subsection{A Specific Design of Proposed MyWear}

\begin{figure} [htbp]
\centering
\subfigure[MyWear Prototype]{
	\includegraphics[height=0.55\textwidth]{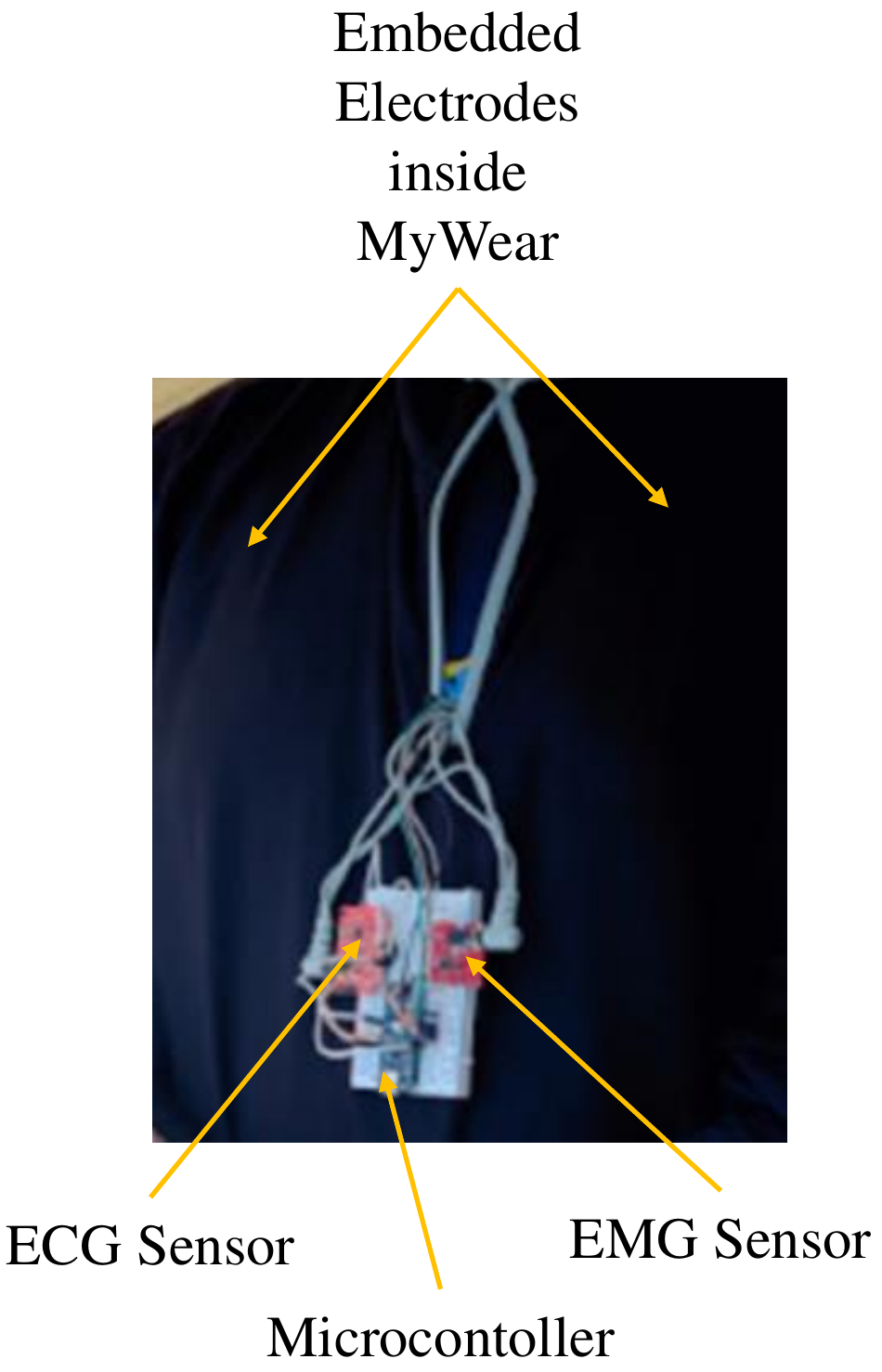}
   	\label{FIG:MyWear_Prototype}
}
\hspace{1cm}
\subfigure[MyWear's mobile application displaying the body vitals]{
	
	\includegraphics[height=0.65\textwidth]{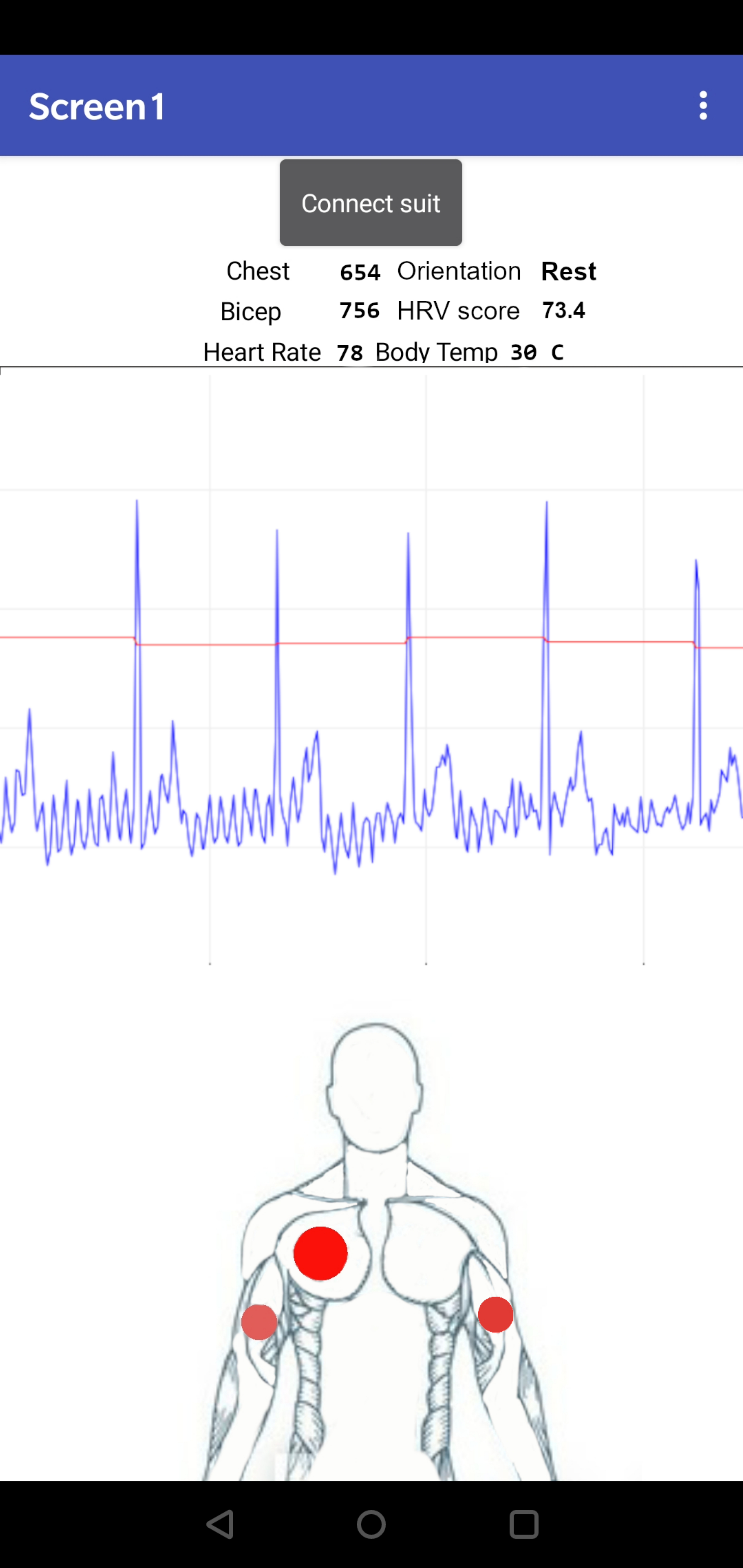}
	\label{FIG:Mobile_App_for_User}
}
\caption{Prototyping of the MyWear using off-the-shelf components.}
\end{figure}

Fig. \ref{FIG:MyWear_Prototype} shows the photograph of the experimental prototype. The characterization table of the MyWear system is given in Table \ref{TBL:characterization_MyWear_System}. The sensors acquire the body vital data upon initialization. The filtered data is sent to the smartphone application using bluetooth. A copy of data is sent to the cloud for analysis and backup through Wi-Fi connectivity.The application displays vitals data as shown in the Fig. \ref{FIG:Mobile_App_for_User}. The temperature and body orientation is displayed. ECG samples of 4 minute time duration are recorded every 20 minutes. In case, no heartbeat is detected, the user is prompted to check whether the electrodes are in contact or not. The HRV score is calculated from ECG as discussed later which helps in determining the stress level of the user. The muscle activity and its intensity is visualized on the human map in forms of colors. The darker the color is, the larger the exerted muscle force is for the particular muscle.

\begin{table}[htbp]
\centering
\caption{Characterization of MyWear System.}
\label{TBL:characterization_MyWear_System}
\begin{tabular}{|p{3cm}|p{6cm}|}
 \hline 
\textbf{Characteristics} & \textbf{Specifics} \\
 \hline
 \hline
Sensor System   &EMG, ECG, IMU, Temperature Sensor\\
\hline
Classifier & Deep Neural Network: C25\\
\hline
Input Dataset & MIT-BIH Arrythmia Dataset\\
\hline
Data Acquisition & Mobile Application and Cloud Service \\
\hline
Connectivity & Bluetooth and Wi-Fi\\
\hline
DNN Accuracy & 98.2\% (worst case) \\
 \hline
\end{tabular}
\end{table}%

The vital data is stored in a Firebase database (cloud platform) that is also capable of running a deep learning model. The model detects whether the heartbeat is normal or irregular. The same is repeated for two samples. If an abnormal or irregular beat is detected, a prompt is then sent to the user's application and an alert is sent to the prescribed doctor and medical officials for immediate assistance.

\subsection{Validation of Detecting Muscle Activity}

\begin{figure}[htbp]
\centering
\subfigure[Muscle Activity -  Bicep]{
	
	\includegraphics[width=0.70\textwidth]{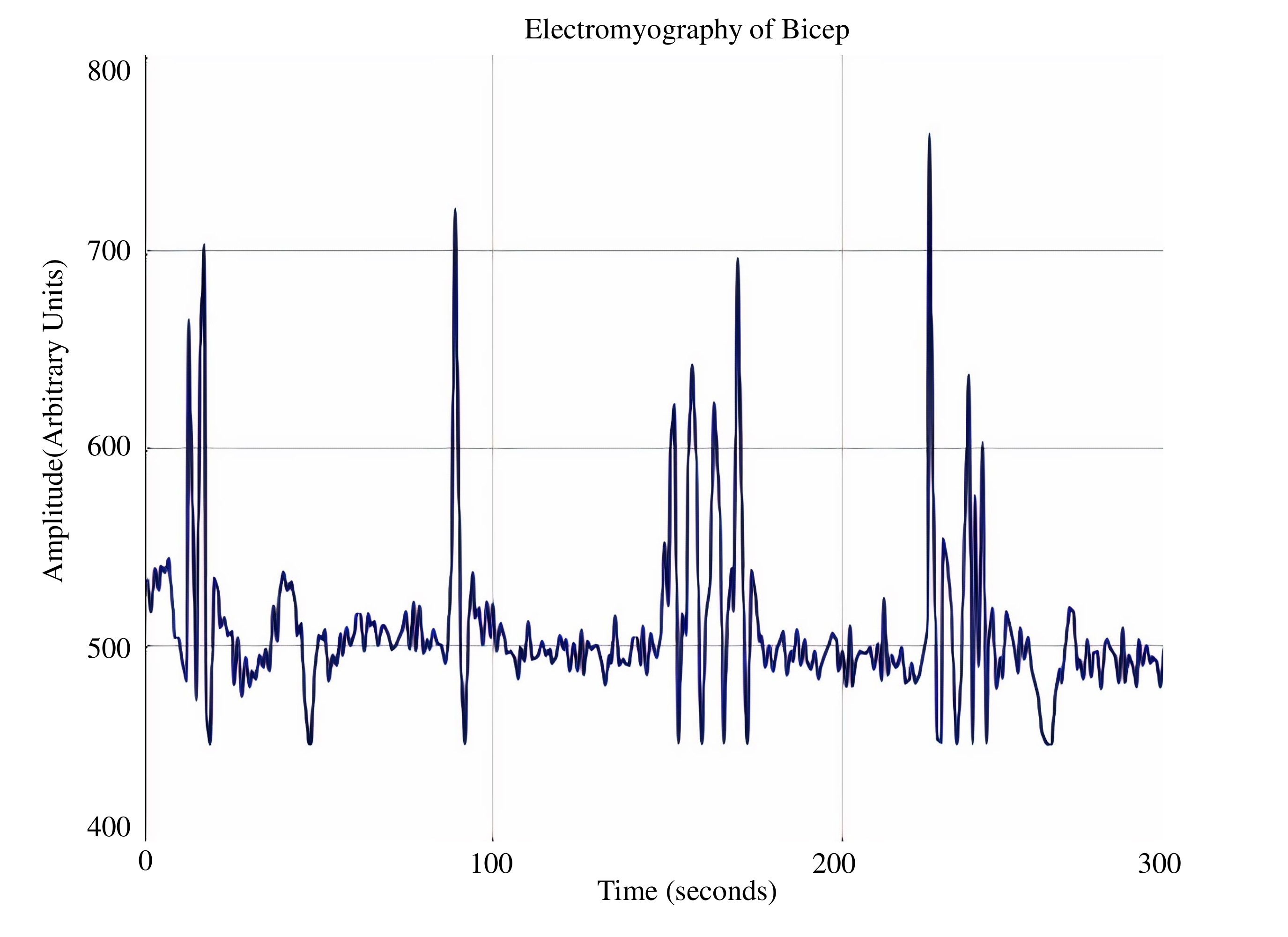}
   	\label{FIG:EMG_Bicep_With_Peaks}
}
\subfigure[Muscle Activity -  Chest]{
	
	\includegraphics[width=0.70\textwidth]{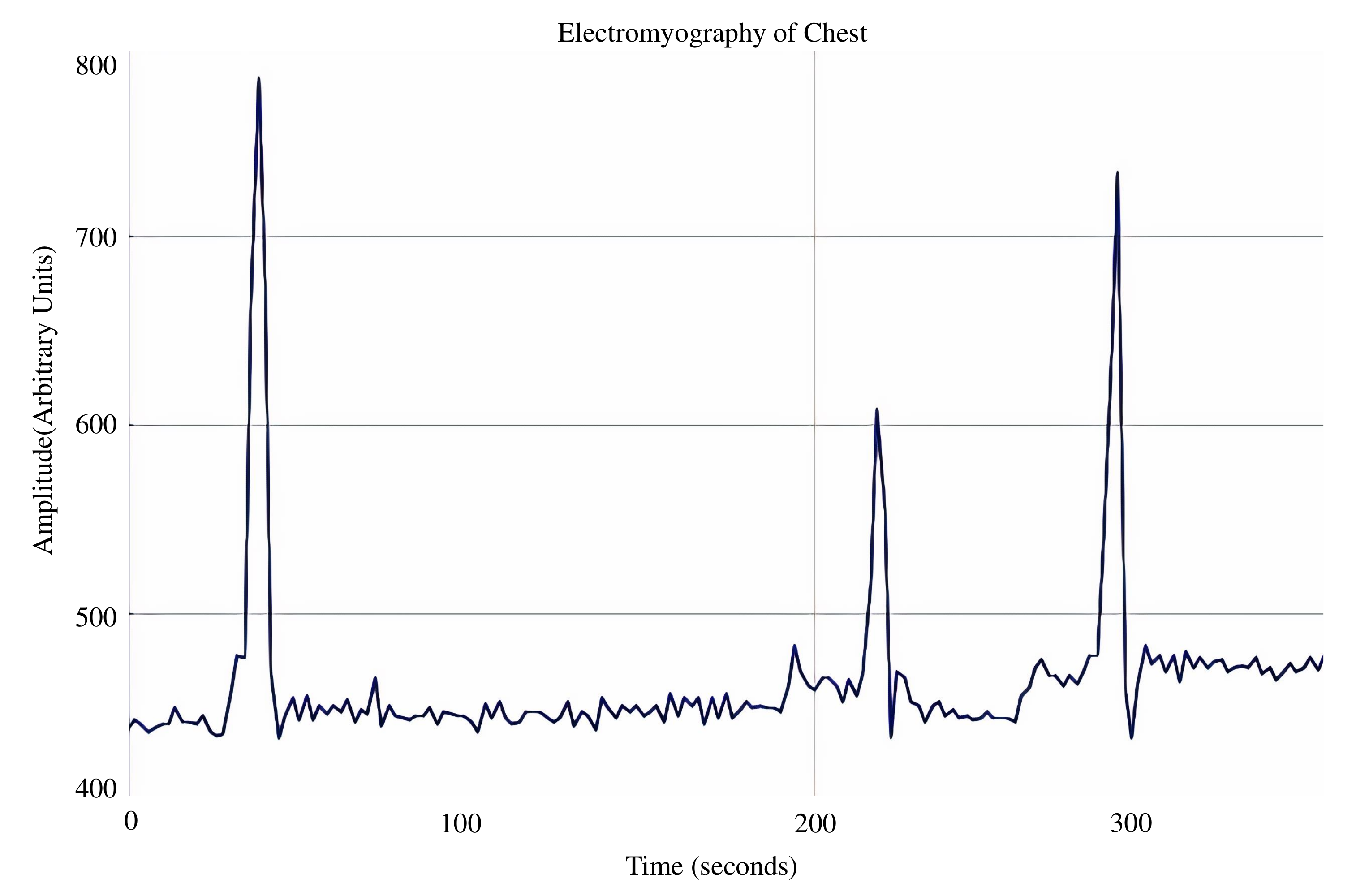}
	\label{FIG:EMG_Chest_With_Peaks}
}
\caption{Muscle Activity detection using EMG with highlighted peaks taken from Bicep and Chest.}
\end{figure}

Overall, 5 tests of 20 minutes were carried out to study the relation of flexing of muscle on the intensity of activity recorded. And the recorded electrical activity was plotted with respect to time. It was noted that while flexing, there was an increase in muscle activity depicted by peaks in the graph. The peaks were sharper and taller when the subject flexed a muscle with greater effort, hence conveying high intensity of muscle force. It was concluded that the higher the intensity the taller the peaks in the graph, suggesting that a greater exert force exerted by the muscle. Fig. \ref{FIG:EMG_Bicep_With_Peaks} and Fig. \ref{FIG:EMG_Chest_With_Peaks} shows the plotted graph depicting instances of increase in muscle intensity in left bicep brachii (bicep) and pectoralis major (chest), respectively.

\subsection{Validating ECG Classification}

\begin{figure} [htbp]
\centering
\subfigure[ECG of a healthy subject]{
	\centering
	\includegraphics[width=0.75\textwidth]{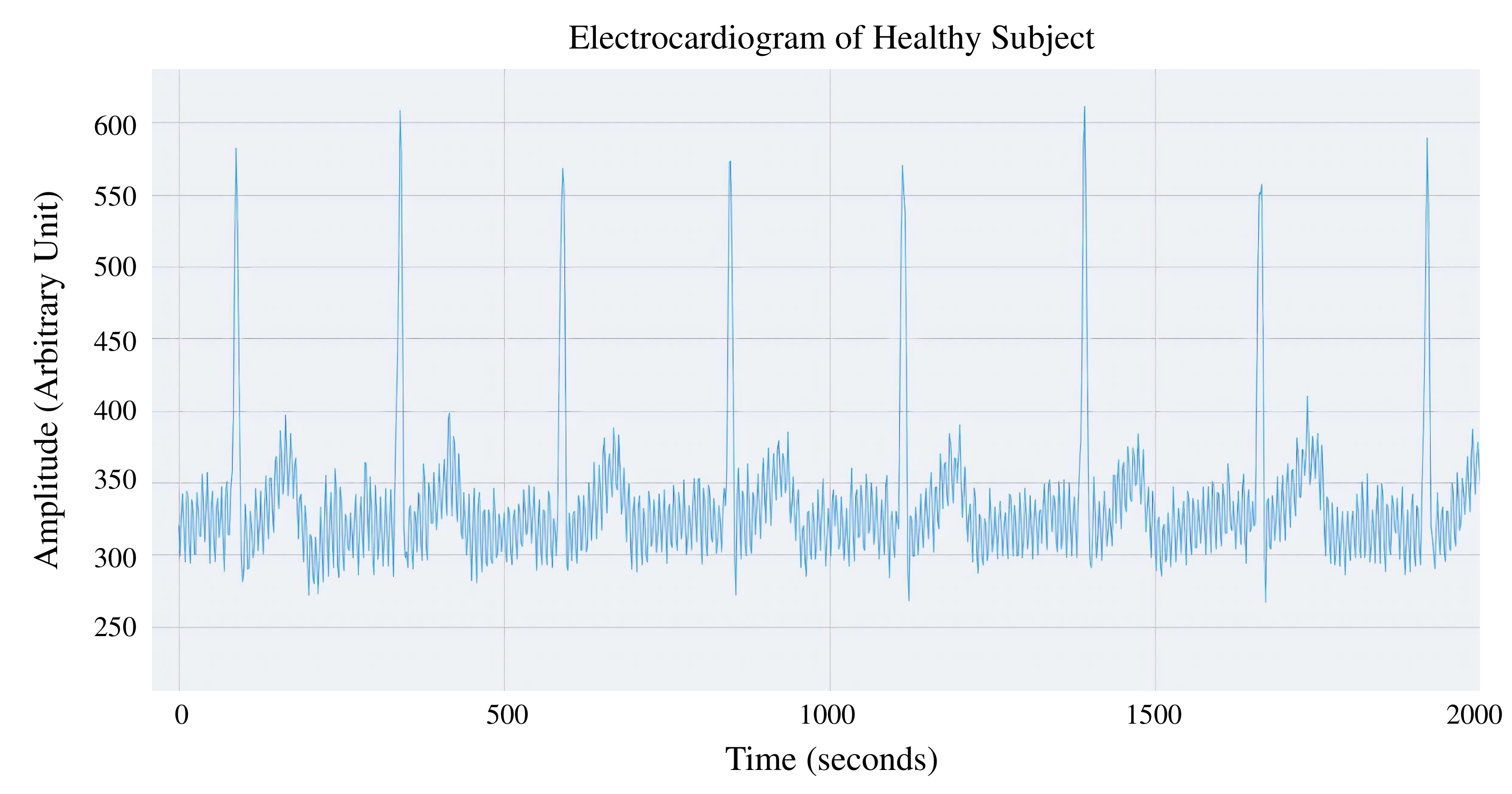}
	\label{FIG:ECG_Healthy_Subject}
}
\subfigure[Normal heartbeat of a healthy subject]{
	\centering
	\includegraphics[width=0.75\textwidth]{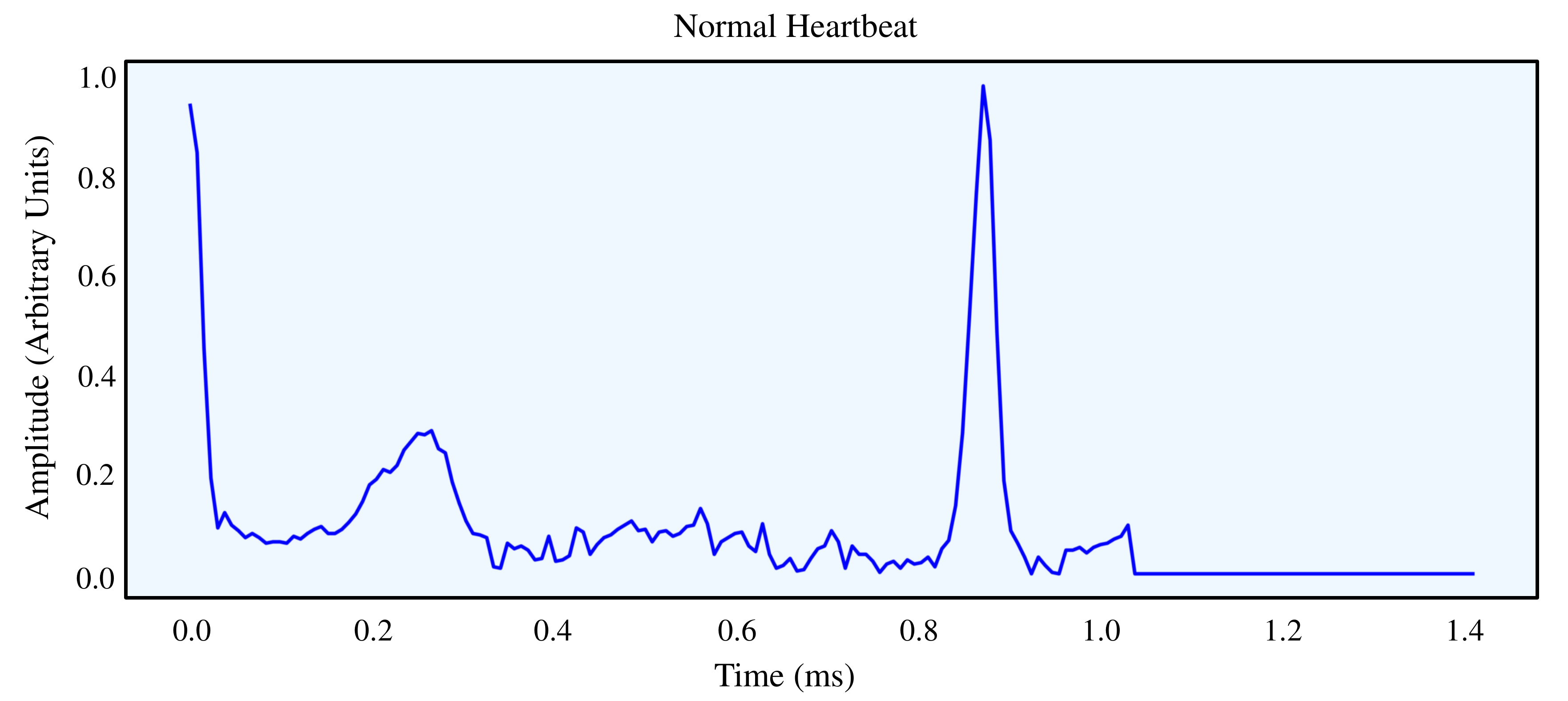}
	\label{FIG:Heartbeat_Normal_Healthy_Person}
}
\caption{ECG and Normal Heartbeat of a healthy subject.}
\end{figure}

Electrocardiogram was collected from three different subjects wearing the garment. ECG was plotted at three different time frames. Fig. \ref{FIG:ECG_Healthy_Subject} shows the Electrocardiogram value of a healthy subject.

\begin{figure*}[htbp]
\centering
\subfigure[Supraventricular]{
	\centering
	\includegraphics[width=0.65\textwidth]{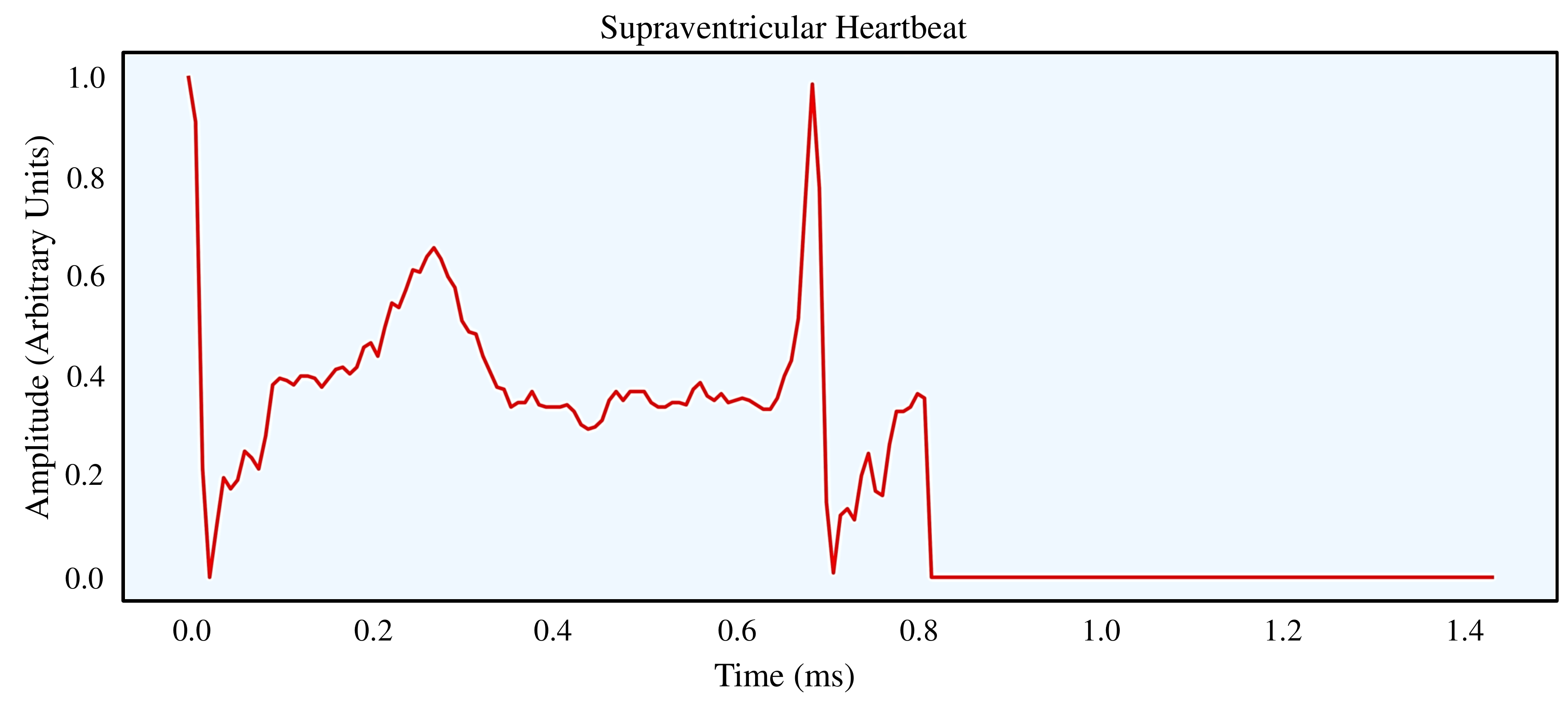}
	\label{FIG:Heartbeat_Abnormal_Case1_Supraventricular}
}
\subfigure[Ventricular]{
	\centering
	\includegraphics[width=0.65\textwidth]{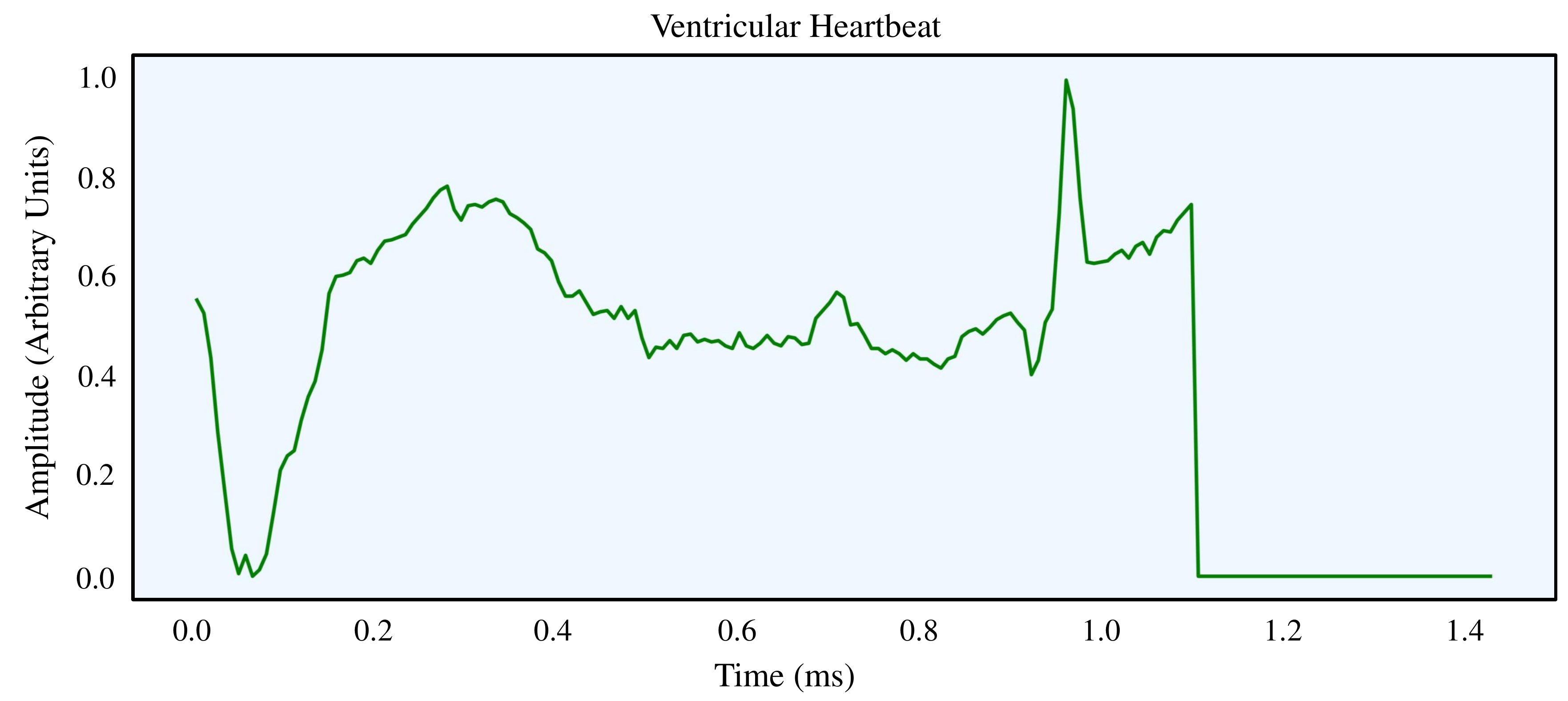}
	\label{FIG:Heartbeat_Abnormal_Case2_Ventricular}
}\\
\subfigure[Fusion]{
	\centering
	\includegraphics[width=0.65\textwidth]{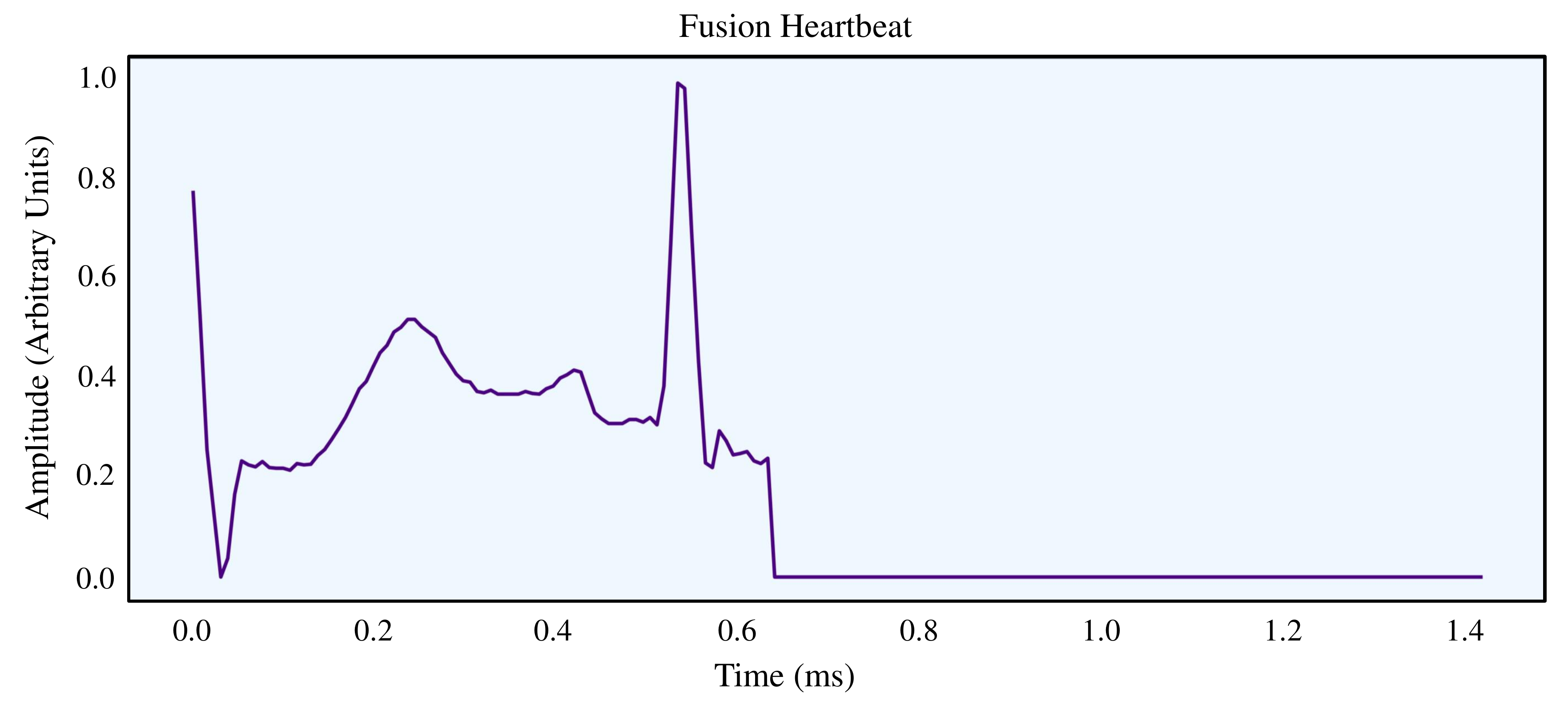}
	\label{FIG:Heartbeat_Abnormal_Case3_Fusion}
}
\subfigure[Unknown]{
	\centering
	\includegraphics[width=0.65\textwidth]{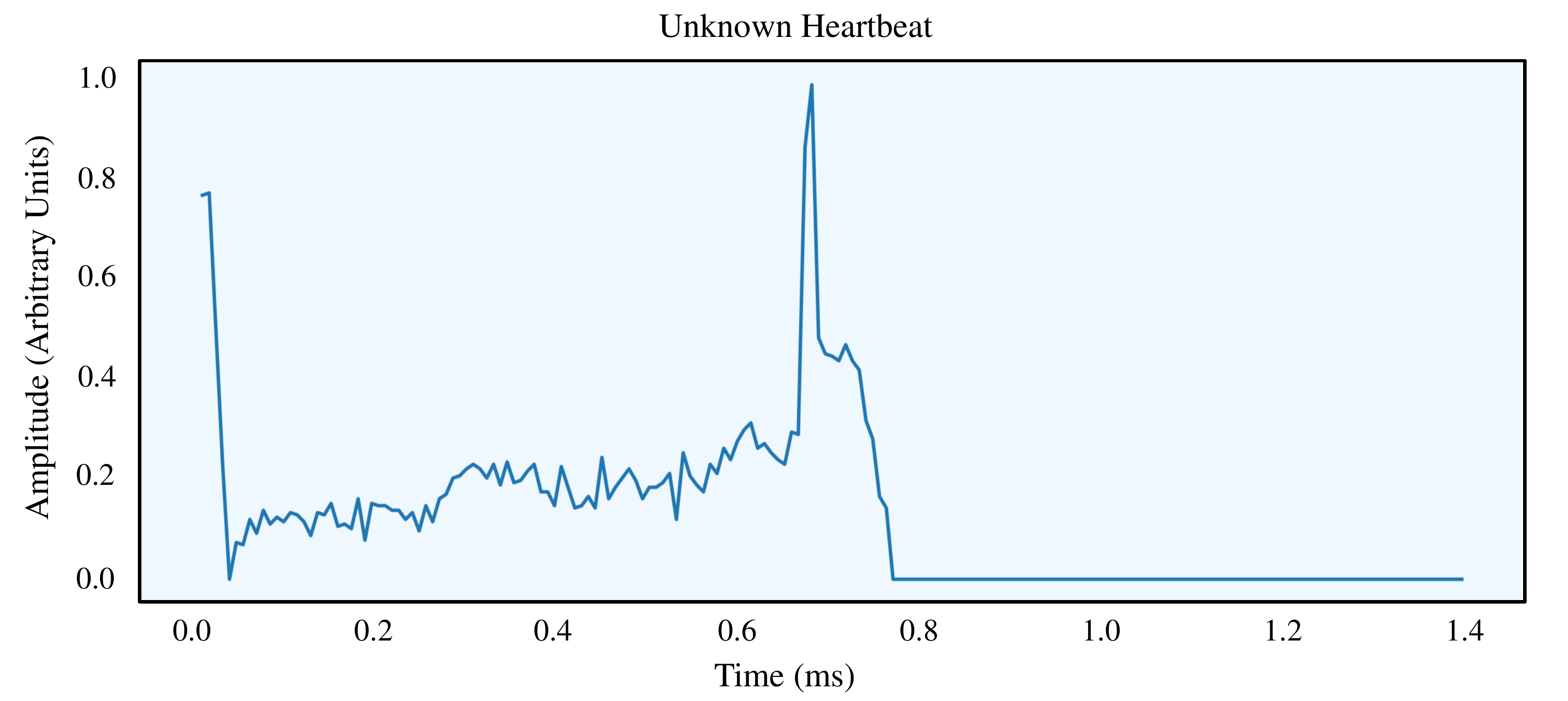}
	\label{FIG:Heartbeat_Abnormal_Case4_Unknown}
}
\caption{Abnormal heartbeats in various forms: a. Supraventricular b. Ventricular  c.  Fusion d. Unknown.}
\label{FIG:Heartbeat_Abnormal_4-Cases}
\end{figure*}

The healthy subject showed a HRV score of 71.87. The HRV score is equal to the RMSSD value as discussed in the existing literature. Fig. \ref{FIG:Heartbeat_Normal_Healthy_Person} depicts a normal heartbeat and Fig. \ref{FIG:Heartbeat_Abnormal_4-Cases} shows different type abnormal heartbeats extracted from MIT-BIH Arrhythmia Dataset \cite{Moody_2001}. Fig. \ref{FIG:Heartbeat_Abnormal_4-Cases} shows the different abnormal heartbeat visualized into the following:
\begin{enumerate}
	\item
Supraventricular 

\item
Unknown

\item
Ventricular

\item
Fusion
\end{enumerate}
The proposed DNN model classifies ECG on the basis of the mentioned type of abnormal heartbeats

\subsection{Validation of Stress (HRV) Detection - Time-Domain Metrics}

\begin{figure}[htbp]
	\centering
	\includegraphics[width=0.45\textheight]{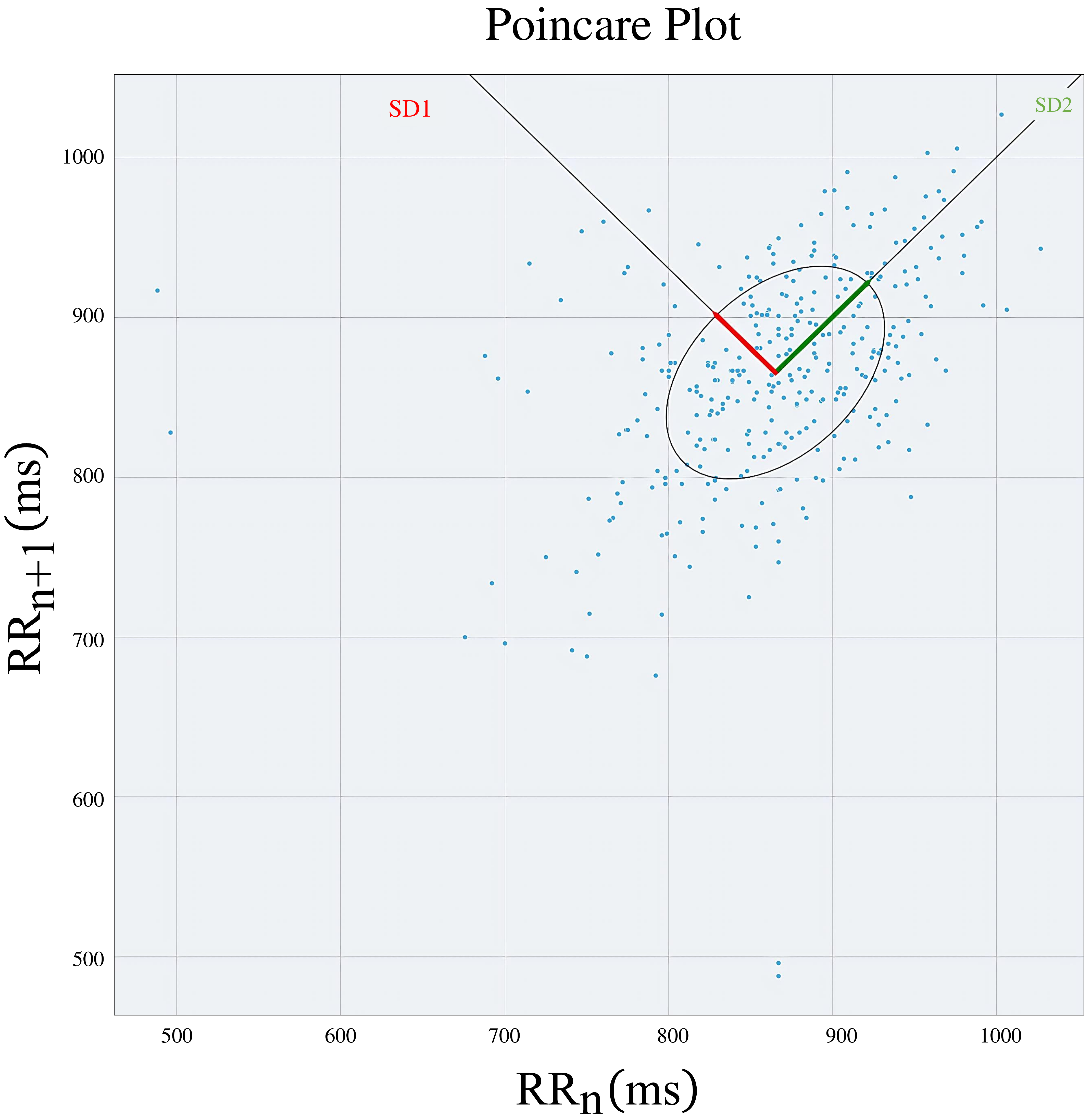}
	\caption{Poincare plot of heart rate variability (HRV).}
	\label{FIG:Results_Poincare-Plot-HRV}
\end{figure}

Table \ref{TBL:Time_domain_metric} shows the time domain metrics obtained after analyzing a 5 minute ECG sample. Fig. \ref{FIG:Results_Poincare-Plot-HRV} shows frequency domain metric Poincare plot. The greater the distance between values in plot, the higher the HRV and lower are the stress levels of the user. Poincare plot is a frequency domain metric or analysis wherein R-R intervals are plotted as a function of the previous RR-interval. The values of each pair of consecutive RR intervals represent a point in the plot. The plot consists of Standard Deviation SD1 and SD2 that correspond to the Standard Deviation of RR interval and Standard Deviation of Successive Difference of RR interval, respectively.

\begin{table}[htbp]
\centering
\caption{Time Domain Metric and Values}
\label{TBL:Time_domain_metric}
\begin{tabular}{|p{3cm}|p{3cm}|}
 \hline 
\textbf{Time Domain Metric} & \textbf{Values} \\
 \hline
 \hline
Mean RR (ms)   &  865.41\\
\hline
STD RR/SDNN (ms) & 66.51\\
\hline
Mean HR (beats/min) & 69.81\\
\hline
STD HR (beats/min) & 6.40 \\
\hline
Min HR (beats/min) & 58.42 \\
\hline
Max HR (beats/min) & 122.95 \\
\hline
RMSSD (ms) & 71.87 \\
\hline
NNxx & 123.00 \\
\hline
pNNxx (\%) & 35.04 \\
 \hline
\end{tabular}
\end{table}%

\subsection{Validation of the DNN Model for for Heart Arrhythmia}

\begin{figure*} [htbp]
\centering
\subfigure[Loss]{
	\includegraphics[width=0.60\textwidth]{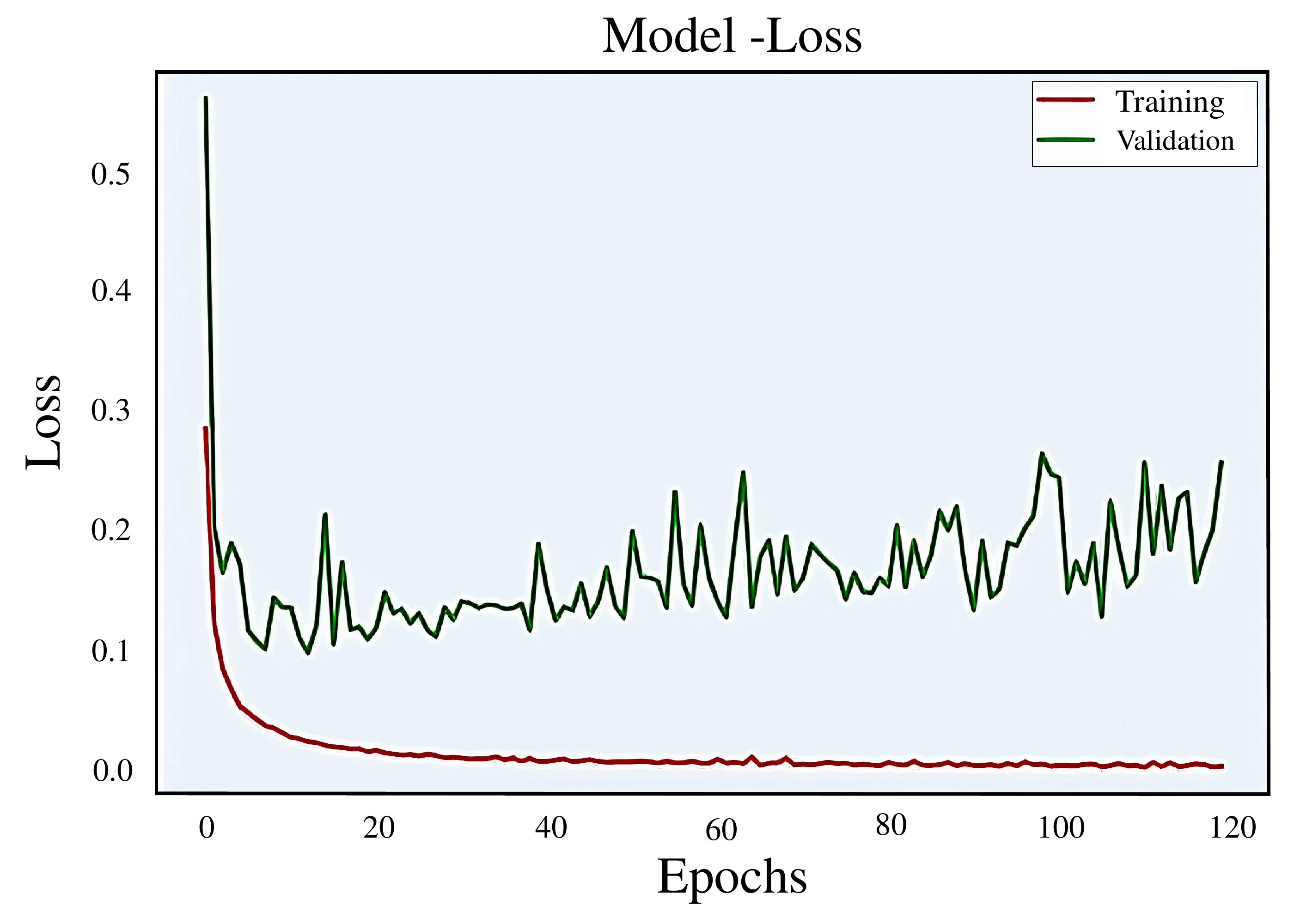}
	\label{FIG:Results_Loss}
}
\subfigure[Accuracy]{
	\centering
	\includegraphics[width=0.60\textwidth]{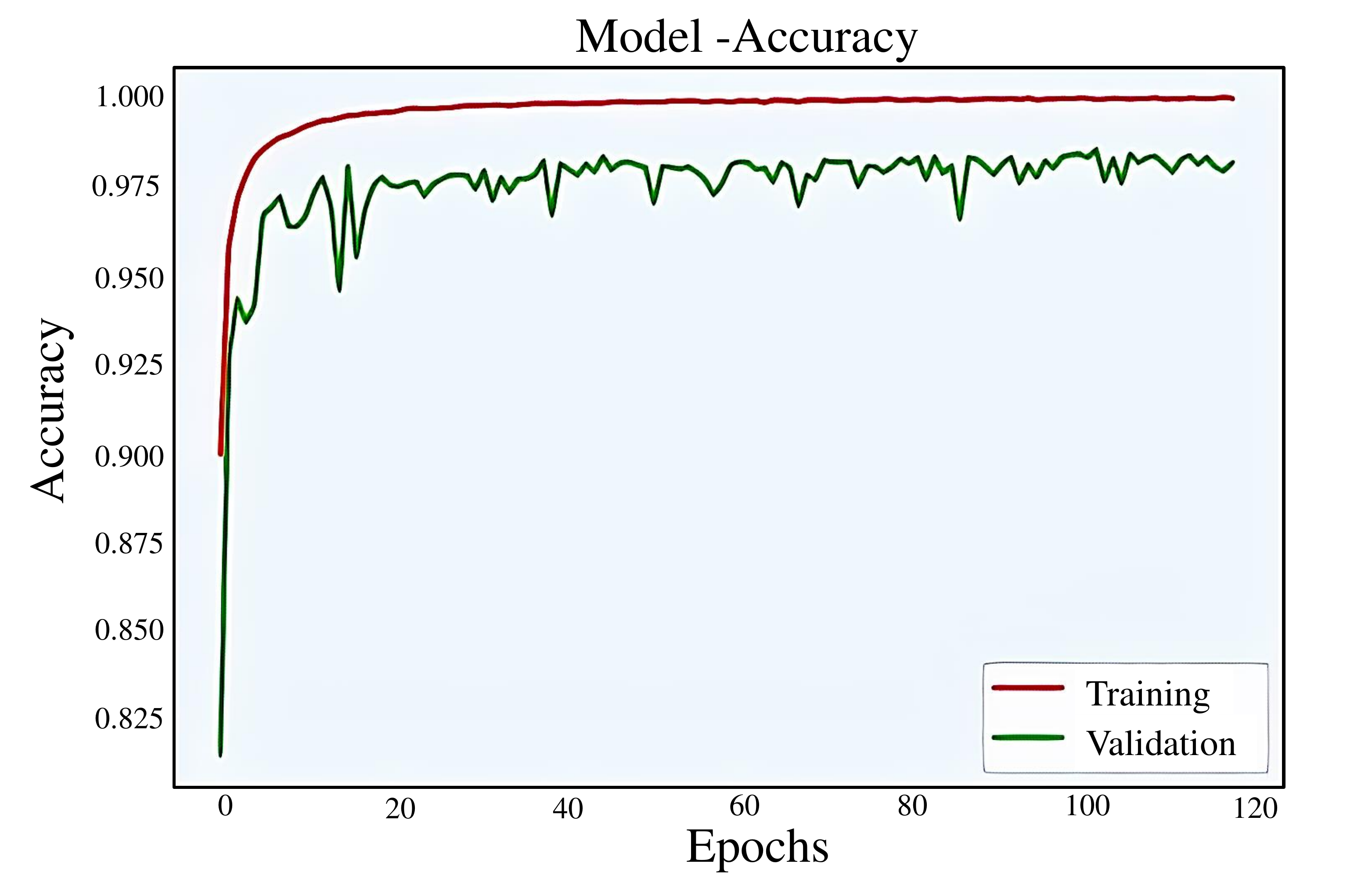}
	\label{FIG:Results_Accuracy}
}
\subfigure[Recall and Precision]{
	\centering
	\includegraphics[width=0.60\textwidth]{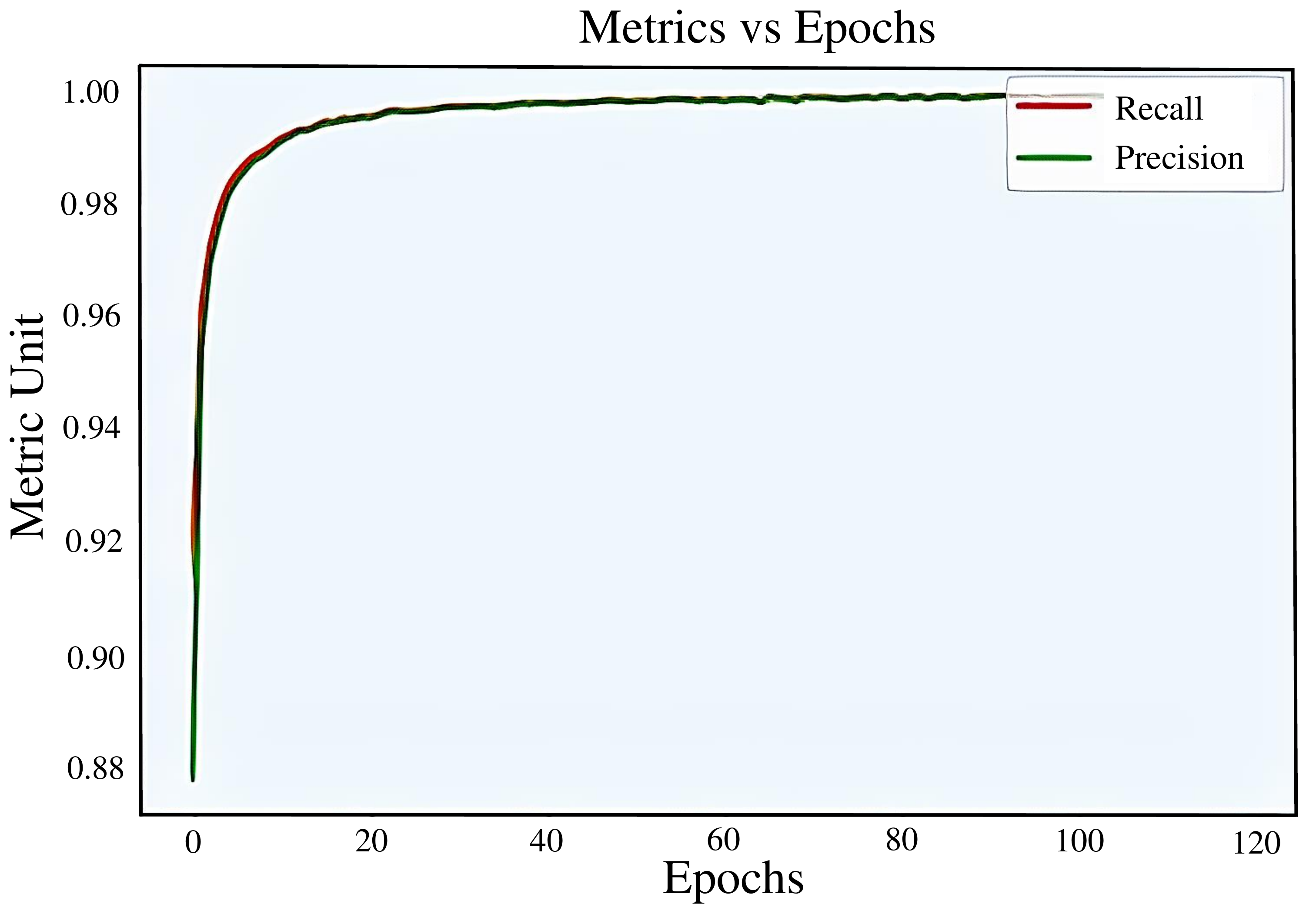}
	\label{FIG:Results_Performance-Recall-Precision}
}
\caption{Performance measures of the proposed DNN model.}
\label{FIG:Performance_measure}
\end{figure*}

The accuracy obtained for the proposed model is 98.2\%. The rate of learning of the model was 0.001. The pattern of maintaining accuracy and tracking loss in classifying heartbeats on the basis of defined categories is shown Fig. \ref{FIG:Results_Accuracy}  and Fig. \ref{FIG:Results_Loss}, respectively. Loss represents the rate at which the model is optimized with respect to the number of epochs the model is trained. The accuracy curve represents the rate at which the system is trained and has reached the maximum level of predicting the heartbeats.

\begin{figure}[htbp]
	\centering
	\includegraphics[width=0.65\textwidth]{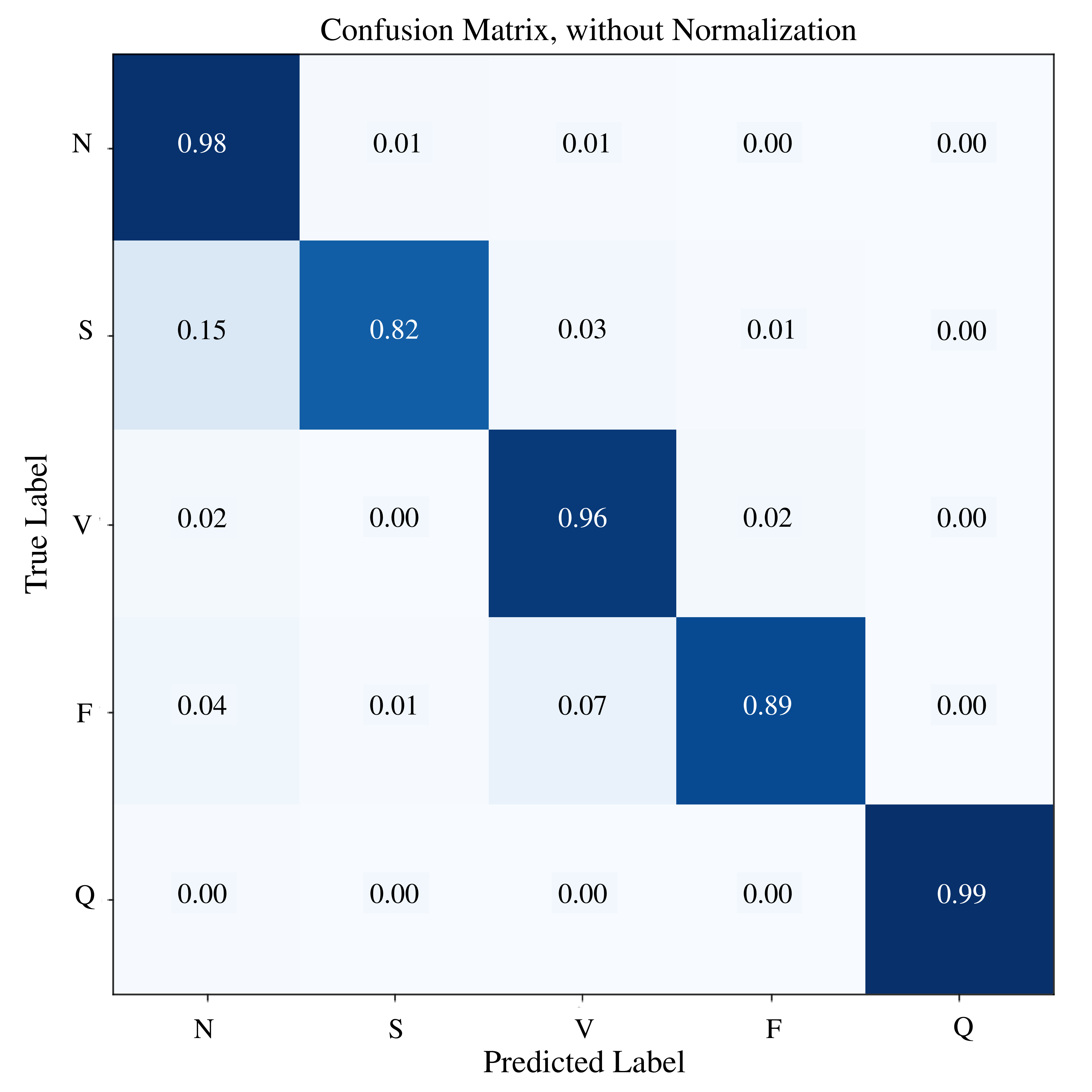}
	\caption{Confusion matrix for DNN heartbeat classification model validation.}
	\label{FIG:Results_Confusion-Matrix}
\end{figure}

Fig. \ref{FIG:Results_Performance-Recall-Precision} shows a plot depicting the relation of performance of recall and precision metric with the number of epochs trained with. The average accuracy of the model is 96.9\% while the precision 97.3\% and recall 97.1\%. 
Table \ref{TBL:Heartbeat_classification_results} shows the comparison of the heartbeat classification results with other state of the art models \cite{Acharya_Raj_Oh_2017, Martis_WSP_2013, Li_Entropy_2016}. Fig. \ref{FIG:Results_Confusion-Matrix} shows the confusion matrix depicting the predicted and truth label after classification of target variables.

\begin{table}[htbp]
\centering
\caption{Comparison of heartbeat classification results.}
\label{TBL:Heartbeat_classification_results}
\begin{tabular}{|p{3.6cm}|p{3.4cm}|p{3.3cm}|}
 \hline 
\textbf{Methodology} & \textbf{Approach} & \textbf{Average Accuracy (\%)} \\
 \hline
 \hline
Raj et al. \cite{Raj_Ray_2018} &DCST + ABC-SVM & 96.1 
\\
\hline
Wang et al. \cite{Wang_TCE_2016-May}  &H-Box  &95 
\\
\hline
Acharya et al. \cite{Acharya_Raj_Oh_2017} &Augmentation + CNN &93.5
\\
\hline
Martis et al. \cite{Martis_WSP_2013}    &DWT + SWM &93.8
\\
\hline
Li et al. \cite{Li_Entropy_2016} & DWT + random forest &94.6
\\
\hline
\textbf{MyWear (Current Paper)}   & DNN    &96.9 \\
 \hline
\end{tabular}
\end{table}%

\begin{table}[htbp]
\centering
\caption{Comparison of classifying myocardial infarction results.}
\label{TBL:Comparison_results_myocardial infarction}
\begin{tabular}{|p{3.5cm}|p{1.9cm}|p{1.9cm}|p{1.5cm}|}
 \hline 
\textbf{Methodology} & \textbf{Accuracy (\%)} & \textbf{Precision (\%)} & \textbf{Recall (\%)} \\
 \hline
 \hline
Raj et al. \cite{Raj_Ray_2018} &96.1 & NA & NA
 \\
\hline
Acharya et al. \cite{Acharya_Rajendra_2017}  &93.5 &92.8 &93.7 
\\
\hline
Kojuri et al. \cite{Kojuri_Javad_2015}    &95.6 &97.9 &93.7 
\\
\hline
Sharma et al. \cite{Sharma_ITBE_2015} &96 &99 &93 
\\
\hline
\textbf{MyWear (Current Paper)}  & 98.2    &97.3  &97.1 
\\
 \hline
\end{tabular}
\end{table}

Table \ref{TBL:Comparison_results_myocardial infarction} shows the comparison of classifying myocardial infarction results with other models \cite{Kojuri_Javad_2015, Sharma_ITBE_2015, Acharya_Raj_Oh_2017}. Fig. \ref{FIG:Mobile_App_for_User} shows the MyWear's mobile application displaying the body vitals such as Heart rate, body temperature, body orientation and HRV score along with plotting ECG data in real-time. The application visualizes the muscle activity on the human map.

\begin{table}[htbp]
\centering
\caption{Comparison of Fall detection results.}
\label{TBL:Comparison_results_Fall_detection}
\begin{tabular}{|p{3.5cm}|p{1.9cm}|p{2.19cm}|p{2.2cm}|}
 \hline 
\textbf{Methodology} & \textbf{Accuracy (\%)} & \textbf{Sensitivity (\%)} & \textbf{Specificity (\%)} \\
 \hline
 \hline
Hemalatha et al. \cite{HEMALATHA_2013} &92 & NA & NA
 \\
 \hline
Mezghani  et al. \cite{Mezghan_2017}  &98 &97.5 &98.5
\\
 \hline
Wang et al. \cite{Wang_TCE_2016-May}    &95 & NA & NA 
\\
 \hline
\textbf{MyWear (Current Paper)}  & 98.5   &98  &99.5 
\\
 \hline
\end{tabular}
\end{table}

\subsection{Validation of Fall Prediction and Detection}

Fig. \ref{FIG:Results_Fall_detection_prediction} depicts the instance at which the model predicts when the person/user is about to experience a fall by recognizing a sudden and quick drop in the resultant acceleration. And a drop is detected when the resultant acceleration drops below $1g$ and quickly increases over $+1g$. Table \ref{TBL:Comparison_results_Fall_detection} shows the comparison of Fall detection results with other models \cite{HEMALATHA_2013, Mezghan_2017, Wang_TCE_2016-May}.

\section{Conclusion and Future Research}
\label{Sec:Conclusion}

Body vital provides insights into the life and lifestyle of the user. Analyzing them provides the user information to improve the health of individuals on a daily basis. Approach presented in this paper helps to enhance the mental and physical state of the user based on analysis of ECG and EMG data, respectively. The proposed garment is integrated with a deep learning model running in IoMT-cloud server that helps in detecting any abnormalities in the heart beat and classifies into the type of abnormality detected. The average accuracy and precision of the proposed deep learning model was 96.9\% and 97.3\%, respectively. MyWear can help in rehabilitation of athletes and sportsmen with the help of embedded sensors that detect Muscle activity and body movement to come up with help for overall body development.

Further, implementing the deep learning model on edge platforms would reduce computational time and resources hence giving results quicker. This can be an extension of the proposed garment and potentially future improvement. First this cane be integrated in IoMT-edge paradigm with TinyML models to rapidly detect the health conditions fast and at the user end \cite{8684800, 9085930, 8922820}. It is a fact that various security and privacy related challenges arise in IoMT driven H-CPS that make smart healthcare. So, blockchain data and device management in of MyWear in IoMT or H-CPS needs serious research \cite{8926457, 8595469, 8662009}.

\bibliographystyle{IEEEtran}

\begin{thebibliography}{10}
	\providecommand{\url}[1]{#1}
	\csname url@samestyle\endcsname
	\providecommand{\newblock}{\relax}
	\providecommand{\bibinfo}[2]{#2}
	\providecommand{\BIBentrySTDinterwordspacing}{\spaceskip=0pt\relax}
	\providecommand{\BIBentryALTinterwordstretchfactor}{4}
	\providecommand{\BIBentryALTinterwordspacing}{\spaceskip=\fontdimen2\font plus
		\BIBentryALTinterwordstretchfactor\fontdimen3\font minus
		\fontdimen4\font\relax}
	\providecommand{\BIBforeignlanguage}[2]{{%
			\expandafter\ifx\csname l@#1\endcsname\relax
			\typeout{** WARNING: IEEEtran.bst: No hyphenation pattern has been}%
			\typeout{** loaded for the language `#1'. Using the pattern for}%
			\typeout{** the default language instead.}%
			\else
			\language=\csname l@#1\endcsname
			\fi
			#2}}
	\providecommand{\BIBdecl}{\relax}
	\BIBdecl
	
	\bibitem{Aazam_MCE_2020-Mar}
	M.~{Aazam}, S.~{Zeadally}, and K.~A. {Harras}, ``Health fog for smart
	healthcare,'' \emph{IEEE Consum. Electron. Mag.}, vol.~9, no.~2, pp. 96--102,
	Mar 2020.
	
	\bibitem{JBHI.2020.2973467}
	H.~{Qiu}, M.~{Qiu}, M.~{Liu}, and G.~{Memmi}, ``Secure health data sharing for
	medical cyber-physical systems for the healthcare 4.0,'' \emph{IEEE Journal
		of Biomedical and Health Informatics}, vol.~24, no.~9, pp. 2499--2505, 2020.
	
	\bibitem{Ghamari_Sensors_2016-Jun}
	M.~Ghamari, B.~Janko, R.~Sherratt, W.~Harwin, R.~Piechockic, and C.~Soltanpur,
	``{A Survey on Wireless Body Area Networks for eHealthcare Systems in
		Residential Environments},'' \emph{MDPI Sensors}, vol.~16, no.~6, p. 831, Jun
	2016.
	
	\bibitem{Joshi_MCE.2020.3018775}
	A.~M. {Joshi}, U.~P. {Shukla}, and S.~P. {Mohanty}, ``Smart healthcare for
	diabetes during {COVID-19},'' \emph{IEEE Consumer Electronics Magazine}, no.
	10.1109/MCE.2020.3018775, pp. 1--1, 2020.
	
	\bibitem{Hsu_MCE_2020-Jan}
	F.~R. {Hsu}, Y.~{Kuo}, S.~{Wei}, Y.~{Hsieh}, and D.~C. {Nguyen}, ``A study of
	user interface with wearable devices based on computer vision,'' \emph{IEEE
		Consumer Electronics Magazine}, vol.~9, no.~1, pp. 43--48, January 2020.
	
	\bibitem{MCE.2020.2969202}
	C.~P. {Antonopoulos}, G.~{Keramidas}, N.~S. {Voros}, M.~{Huebner},
	F.~{Schwiegelshohn}, D.~{Goehringer}, M.~{Dagioglou}, G.~{Stavrinos},
	S.~{Konstantopoulos}, and V.~{Karkaletsis}, ``Toward an {ICT}-based service
	oriented health care paradigm,'' \emph{IEEE Consumer Electronics Magazine},
	vol.~9, no.~4, pp. 77--82, 2020.
	
	\bibitem{thomson_nuss_2019}
	E.~A. Thomson, K.~Nuss, A.~Comstock, S.~Reinwald, S.~Blake, R.~E. Pimentel,
	B.~L. Tracy, and K.~Li, ``Heart rate measures from the apple watch, fitbit
	charge hr 2, and electrocardiogram across different exercise intensities,''
	\emph{J. of Sports Sci.}, vol.~37, no.~12, p. 1411–1419, 2019.
	
	\bibitem{MCE.2019.2956205}
	A.~{Petropoulos}, D.~{Sikeridis}, and T.~{Antonakopoulos}, ``Wearable smart
	health advisors: An {IMU}-enabled posture monitor,'' \emph{IEEE Consumer
		Electronics Magazine}, vol.~9, no.~5, pp. 20--27, 2020.
	
	\bibitem{Zhu_MCE_2019-Sep}
	H.~{Zhu}, C.~K. {Wu}, C.~H. {Koo}, Y.~T. {Tsang}, Y.~{Liu}, H.~R. {Chi}, and
	K.~{Tsang}, ``{Smart Healthcare in the Era of Internet-of-Things},''
	\emph{IEEE Consum. Electron. Mag.}, vol.~8, no.~5, pp. 26--30, Sep 2019.
	
	\bibitem{Gonzales_2019}
	\BIBentryALTinterwordspacing
	M.~Gonzales, ``Fitness trackers: How they work and their highly anticipated
	future,'' \emph{Health \& Medicine}, vol. XIX, no.~II, April 2019. [Online].
	Available:
	\url{https://illumin.usc.edu/fitness-trackers-how-they-work-and-their-highly-anticipated-future/}
	\BIBentrySTDinterwordspacing
	
	\bibitem{Joshi_TCE.2020.3011966}
	A.~M. {Joshi}, P.~{Jain}, S.~P. {Mohanty}, and N.~{Agrawal}, ``{iGLU 2.0}: A
	new wearable for accurate non-invasive continuous serum glucose measurement
	in iomt framework,'' \emph{IEEE Transactions on Consumer Electronics}, no.
	10.1109/TCE.2020.3011966, pp. 1--1, 2020.
	
	\bibitem{Jain_MCE_2020-Jan}
	P.~{Jain}, A.~M. {Joshi}, and S.~P. {Mohanty}, ``{iGLU}: An intelligent device
	for accurate noninvasive blood glucose-level monitoring in smart
	healthcare,'' \emph{IEEE Consumer Electronics Magazine}, vol.~9, no.~1, pp.
	35--42, January 2020.
	
	\bibitem{1932296818768618}
	R.~Basatneh, B.~Najafi, and D.~G. Armstrong, ``Health sensors, smart home
	devices, and the internet of medical things: An opportunity for dramatic
	improvement in care for the lower extremity complications of diabetes,''
	\emph{Journal of Diabetes Science and Technology}, vol.~12, no.~3, pp.
	577--586, 2018.
	
	\bibitem{Pandian_2008}
	P.~Pandian, K.~Mohanavelu, K.~Safeer, T.~Kotresh, D.~Shakunthala, P.~Gopal, and
	V.~Padaki, ``Smart {Vest}: {Wearable} multi-parameter remote physiological
	monitoring system,'' \emph{Med. Eng. \& Phy.}, vol.~30, no.~4, pp. 466--477,
	May 2008.
	
	\bibitem{JSEN.2019.2949608}
	C.~{Massaroni}, J.~{Di Tocco}, M.~{Bravi}, A.~{Carnevale}, D.~{Lo Presti},
	R.~{Sabbadini}, S.~{Miccinilli}, S.~{Sterzi}, D.~{Formica}, and E.~{Schena},
	``Respiratory monitoring during physical activities with a multi-sensor smart
	garment and related algorithms,'' \emph{IEEE Sensors Journal}, vol.~20,
	no.~4, pp. 2173--2180, 2020.
	
	\bibitem{JERM.2019.2929676}
	Y.~{Jiang}, K.~{Pan}, T.~{Leng}, and Z.~{Hu}, ``Smart textile integrated
	wireless powered near field communication body temperature and sweat sensing
	system,'' \emph{IEEE Journal of Electromagnetics, RF and Microwaves in
		Medicine and Biology}, vol.~4, no.~3, pp. 164--170, 2020.
	
	\bibitem{Lee_Young_2009}
	Y.-D. Lee and W.-Y. Chung, ``Wireless sensor network based wearable smart shirt
	for ubiquitous health and activity monitoring,'' \emph{Sens. \& Actu. B:
		Chem.}, vol. 140, no.~2, pp. 390--395, Jul 2009.
	
	\bibitem{Rachakonda_TCE_2019}
	L.~{Rachakonda}, S.~P. {Mohanty}, E.~{Kougianos}, and P.~{Sundaravadivel},
	``{Stress-Lysis}: A {DNN}-integrated edge device for stress level detection
	in the {IoMT},'' \emph{IEEE Trans. Consum. Electron.}, vol.~65, no.~4, pp.
	474--483, Nov 2019.
	
	\bibitem{Rachakonda_TCE_2020-May}
	L.~{Rachakonda}, S.~P. {Mohanty}, and E.~{Kougianos}, ``{iLog}: An intelligent
	device for automatic food intake monitoring and stress detection in the
	{IoMT},'' \emph{{IEEE Trans. Consum. Electron.}}, vol.~2, no.~66, pp.
	115--124, May 2020.
	
	\bibitem{MCE.2018.2797741}
	C.~{Lee}, P.~{Chondro}, S.~{Ruan}, O.~{Christen}, and E.~{Naroska}, ``Improving
	mobility for the visually impaired: A wearable indoor positioning system
	based on visual markers,'' \emph{IEEE Consumer Electronics Magazine}, vol.~7,
	no.~3, pp. 12--20, 2018.
	
	\bibitem{Ava_smart_bracelet}
	\BIBentryALTinterwordspacing
	AvaWomen, ``\BIBforeignlanguage{en}{Category: {Women's Health}},'' last
	accessed on 14 Aug 2020. [Online]. Available: \url{https://www.avawomen.com/}
	\BIBentrySTDinterwordspacing
	
	\bibitem{Lin_Lai_2018}
	Y.~{Lin}, Y.~{Lai}, H.~{Chang}, Y.~{Tsao}, Y.~{Chang}, and R.~Y. {Chang},
	``Smarthear: A smartphone-based remote microphone hearing assistive system
	using wireless technologies,'' \emph{IEEE Syst. J.}, vol.~12, no.~1, pp.
	20--29, 2018.
	
	\bibitem{Foroughi_2016}
	J.~{Foroughi}, T.~{Mitew}, P.~{Ogunbona}, R.~{Raad}, and F.~{Safaei}, ``Smart
	fabrics and networked clothing: Recent developments in {CNT}-based fibers and
	their continual refinement,'' \emph{IEEE Consum. Electron. Mag.}, vol.~5,
	no.~4, pp. 105--111, Jul 2016.
	
	\bibitem{Puranik_TCE_2020-Jan}
	S.~{Puranik} and A.~W. {Morales}, ``Heart rate estimation of {PPG} signals with
	simultaneous accelerometry using adaptive neural network filtering,''
	\emph{IEEE Trans. Consum. Electron.}, vol.~66, no.~1, pp. 69--76, Jan 2020.
	
	\bibitem{Kim_TCE_2019-Aug}
	J.~W. {Kim}, J.~H. {Lim}, S.~M. {Moon}, and B.~{Jang}, ``Collecting health
	lifelog data from smartwatch users in a privacy-preserving manner,''
	\emph{IEEE Trans. Consum. Electron.}, vol.~65, no.~3, pp. 369--378, Aug 2019.
	
	\bibitem{Raj_Ray_2018}
	S.~{Raj} and K.~C. {Ray}, ``A personalized point-of-care platform for real-time
	{ECG} monitoring,'' \emph{IEEE Trans. Consum. Electron.}, vol.~64, no.~4, pp.
	452--460, Nov 2018.
	
	\bibitem{Prabha_Saraju_2018}
	P.~{Sundaravadivel}, K.~{Kesavan}, L.~{Kesavan}, S.~P. {Mohanty}, and
	E.~{Kougianos}, ``{Smart-Log}: A deep-learning based automated nutrition
	monitoring system in the iot,'' \emph{IEEE Trans. Consum. Electron.},
	vol.~64, no.~3, pp. 390--398, Sep 2018.
	
	\bibitem{Wang_TCE_2016-May}
	L.~{Wang}, Y.~{Hsiao}, X.~{Xie}, and S.~{Lee}, ``An outdoor intelligent
	healthcare monitoring device for the elderly,'' \emph{IEEE Trans. Consum.
		Electron.}, vol.~62, no.~2, pp. 128--135, May 2016.
	
	\bibitem{Hexoskin_2020}
	\BIBentryALTinterwordspacing
	Hexoskin), ``\BIBforeignlanguage{en}{Hexoskin {Smart} {Shirts} - {Cardiac},
		{Respiratory}, {Sleep} \& {Activity} {Metrics}},'' last accessed on 14 Aug
	2020. [Online]. Available: \url{https://www.hexoskin.com/}
	\BIBentrySTDinterwordspacing
	
	\bibitem{Athos_2018}
	\BIBentryALTinterwordspacing
	Athos, ``\BIBforeignlanguage{en}{Athos {Coaching} {System}},'' last accessed on
	14 Aug 2020. [Online]. Available: \url{https://shop.liveathos.com/}
	\BIBentrySTDinterwordspacing
	
	\bibitem{Farjadian_ICORR_2013}
	A.~B. Farjadian, M.~L. Sivak, and C.~Mavroidis, ``{SQUID}: {Sensorized} shirt
	with smartphone interface for exercise monitoring and home rehabilitation,''
	in \emph{Proc. IEEE Int Conf Rehabil Robot}, 2013, pp. 1--6.
	
	\bibitem{MCE.2017.2776462}
	S.~S. {Roy}, D.~{Puthal}, S.~{Sharma}, S.~P. {Mohanty}, and A.~Y. {Zomaya},
	``Building a sustainable internet of things: Energy-efficient routing using
	low-power sensors will meet the need,'' \emph{IEEE Consumer Electronics
		Magazine}, vol.~7, no.~2, pp. 42--49, March 2018.
	
	\bibitem{MCE.2017.2714695}
	C.~{Yang}, D.~{Puthal}, S.~P. {Mohanty}, and E.~{Kougianos}, ``Big-sensing-data
	curation for the cloud is coming: A promise of scalable cloud-data-center
	mitigation for next-generation {IoT} and wireless sensor networks,''
	\emph{IEEE Consumer Electronics Magazine}, vol.~6, no.~4, pp. 48--56, October
	2017.
	
	\bibitem{8719325}
	L.~{Rachakonda}, P.~{Sundaravadivel}, S.~P. {Mohanty}, E.~{Kougianos}, and
	M.~{Ganapathiraju}, ``A smart sensor in the {IoMT} for stress level
	detection,'' in \emph{Proc. IEEE International Symposium on Smart Electronic
		Systems (iSES) (Formerly iNiS)}, 2018, pp. 141--145.
	
	\bibitem{Mohanty_VAIBHAV_2020_Panel}
	\BIBentryALTinterwordspacing
	S.~P. Mohanty, ``{TinyML} - key for smart cities and smart villages,''
	{VAIBHAV} Summit 2020 Panel -- Vertical: V6 - Data Sciences, Last accessed on
	17 Oct 2020. [Online]. Available:
	\url{http://www.smohanty.org/Presentations/2020/Mohanty_VAIBHAV_2020_V6-Data-Sciences_Session-V6H5S2_Data-Applications.pdf}
	\BIBentrySTDinterwordspacing
	
	\bibitem{8684800}
	D.~{Puthal}, S.~P. {Mohanty}, S.~A. {Bhavake}, G.~{Morgan}, and R.~{Ranjan},
	``Fog computing security challenges and future directions,'' \emph{IEEE
		Consumer Electronics Magazine}, vol.~8, no.~3, pp. 92--96, May 2019.
	
	\bibitem{Iskandar_AISP_2019}
	W.~J. Iskandar, I.~Roihan, and R.~A. Koestoer, ``Prototype low-cost portable
	electrocardiogram ({ECG}) based on {Arduino}-{Uno} with {Bluetooth}
	feature,'' in \emph{AIP Conf. Proc.}, 2019, pp. 050\,019--7.
	
	\bibitem{Kanani_2018}
	P.~Kanani and M.~Padole, ``Recognizing {Real} {Time} {ECG} {Anomalies} {Using}
	{Arduino}, {AD8232} and {Java},'' in \emph{Adv. Comp. and Data Scienc.},
	2018, vol. 905, pp. 54--64.
	
	\bibitem{Villegas_2019}
	A.~Villegas, D.~McEneaney, and O.~Escalona, ``Arm-{ECG} {Wireless} {Sensor}
	{System} for {Wearable} {Long}-{Term} {Surveillance} of {Heart}
	{Arrhythmias},'' \emph{Electronics}, vol.~8, no.~11, p. 1300, Nov 2019.
	
	\bibitem{Samarawickrama_2018}
	K.~Samarawickrama, S.~Ranasinghe, Y.~Wickramasinghe, W.~Mallehevidana,
	V.~Marasinghe, and K.~Wijesinghe, ``Surface emg signal acquisition analysis
	and classification for the operation of a prosthetic limb,'' \emph{Int. J.
		Biosci., Biochem. \& Bioinfo.}, vol.~8, pp. 32--41, 01 2018.
	
	\bibitem{Alimam_2017}
	H.~Alimam, N.~Abo~Alzahab, and M.~Alkhayat, ``Design of emg acquisition circuit
	to control an antagonistic mechanism actuated by pneumatic artificial muscles
	pams,'' \emph{Int. J. Mech. \& Mechatron. Eng.}, vol.~17, pp. 37--47, 10
	2017.
	
	\bibitem{Wu_IS3C_2016}
	M.~{Wu}, S.~{Shieh}, Y.~{Liao}, and Y.~{Chen}, ``{ECG} measurement system based
	on arduino and android devices,'' in \emph{Proc. Int. Sympo. Comp., Consum.
		and Control}, 2016, pp. 690--693.
	
	\bibitem{Wang_mse_2012}
	H.-M. Wang and S.-C. Huang, ``{SDNN}/{RMSSD} as a {Surrogate} for {LF}/{HF}:
	{A} {Revised} {Investigation},'' \emph{Model. \& Simu. in Eng.}, vol. 2012,
	pp. 1--8, 2012.
	
	\bibitem{mindful_HRV_2020}
	\BIBentryALTinterwordspacing
	Mindful, ``\BIBforeignlanguage{en}{Category: {HRV}},'' last accessed on 14 Aug
	2020. [Online]. Available: \url{http://www.mindfulwellness.us/3/category/hrv}
	\BIBentrySTDinterwordspacing
	
	\bibitem{Xiangrui_2020}
	X.~Li, X.~Li, D.~Pan, and D.~Zhu, ``On the learning property of logistic and
	softmax losses for deep neural networks,'' \emph{arXiv}, 2020.
	
	\bibitem{Moody_2001}
	G.~Moody and R.~Mark, ``The impact of the {MIT}-{BIH} {Arrhythmia}
	{Database},'' \emph{IEEE Eng. Med and Bio. Mag.}, vol.~20, no.~3, pp. 45--50,
	Jun. 2001.
	
	\bibitem{husstech_2020}
	\BIBentryALTinterwordspacing
	``Using the accelerometer,'' last accessed on 31 Aug 2020. [Online]. Available:
	\url{http://husstechlabs.com/projects/atb1/using-the-accelerometer/}
	\BIBentrySTDinterwordspacing
	
	\bibitem{Zhou_2018}
	X.~{Zhou}, L.~{Qian}, P.~{You}, Z.~{Ding}, and Y.~{Han}, ``Fall detection using
	convolutional neural network with multi-sensor fusion,'' in \emph{Proc. IEEE
		Int Conf Multi. Expo Workshops (ICMEW)}, 2018, pp. 1--5.
	
	\bibitem{Fan_2020}
	Y.~{Fan}, H.~{Jin}, Y.~{Ge}, and N.~{Wang}, ``Wearable motion attitude
	detection and data analysis based on internet of things,'' \emph{IEEE
		Access}, vol.~8, pp. 1327--1338, 2020.
	
	\bibitem{Mezghan_2017}
	N.~{Mezghani}, Y.~{Ouakrim}, M.~R. {Islam}, R.~{Yared}, and B.~{Abdulrazak},
	``Context aware adaptable approach for fall detection bases on smart
	textile,'' in \emph{Proc. IEEE EMBS Int Conf Bio. Health Info. (BHI)}, 2017,
	pp. 473--476.
	
	\bibitem{HEMALATHA_2013}
	S.~Hemalatha and V.~Vijayakumar, ``Frequent bit pattern mining over tri-axial
	accelerometer data streams for recognizing human activities and detecting
	fall,'' \emph{Procedia Computer Science}, vol.~19, pp. 56 -- 63, 2013.
	
	\bibitem{Acharya_Raj_Oh_2017}
	U.~R. Acharya, S.~L. Oh, Y.~Hagiwara, J.~H. Tan, M.~Adam, A.~Gertych, and R.~S.
	Tan, ``A deep convolutional neural network model to classify heartbeats,''
	\emph{Comp. in Bio. \& Med.}, vol.~89, pp. 389--396, Oct 2017.
	
	\bibitem{Martis_WSP_2013}
	R.~J. Martis, U.~R. Acharya, C.~M. Lim, K.~M. Mandana, A.~K. Ray, and
	C.~Chakraborty, ``Application of higher order cumulant features for cardiac
	health diagnosis using {ECG} signals,'' \emph{Int. J. Neural Sys.}, vol.~23,
	p. 1350014, Aug 2013.
	
	\bibitem{Li_Entropy_2016}
	T.~Li and M.~Zhou, ``{ECG} {Classification} {Using} {Wavelet} {Packet}
	{Entropy} and {Random} {Forests},'' \emph{Entropy}, vol.~18, p. 285, Aug
	2016.
	
	\bibitem{Acharya_Rajendra_2017}
	U.~R. Acharya, H.~Fujita, S.~L. Oh, Y.~Hagiwara, J.~H. Tan, and M.~Adam,
	``Application of deep convolutional neural network for automated detection of
	myocardial infarction using {ECG} signals,'' \emph{Info. Sci.}, vol. 415-416,
	pp. 190--198, Nov 2017.
	
	\bibitem{Kojuri_Javad_2015}
	J.~Kojuri, R.~Boostani, P.~Dehghani, F.~Nowroozipour, and N.~Saki, ``Prediction
	of acute myocardial infarction with artificial neural networks in patients
	with nondiagnostic electrocardiogram,'' \emph{J. Cardiovasc. Dis. Res.},
	vol.~6, no.~2, pp. 51--59, May 2015.
	
	\bibitem{Sharma_ITBE_2015}
	L.~N. {Sharma}, R.~K. {Tripathy}, and S.~{Dandapat}, ``Multiscale energy and
	eigenspace approach to detection and localization of myocardial infarction,''
	\emph{IEEE Trans. Biomed Eng.}, vol.~62, no.~7, pp. 1827--1837, 2015.
	
	\bibitem{9085930}
	A.~K. {Tripathy}, A.~G. {Mohapatra}, S.~P. {Mohanty}, E.~{Kougianos}, A.~M.
	{Joshi}, and G.~{Das}, ``{EasyBand}: A wearable for safety-aware mobility
	during pandemic outbreak,'' \emph{IEEE Consumer Electronics Magazine},
	vol.~9, no.~5, pp. 57--61, 2020.
	
	\bibitem{8922820}
	P.~{Jain}, A.~M. {Joshi}, and S.~P. {Mohanty}, ``{iGLU}: An intelligent device
	for accurate noninvasive blood glucose-level monitoring in smart
	healthcare,'' \emph{IEEE Consumer Electronics Magazine}, vol.~9, no.~1, pp.
	35--42, 2020.
	
	\bibitem{8926457}
	S.~{Biswas}, K.~{Sharif}, F.~{Li}, S.~{Maharjan}, S.~P. {Mohanty}, and
	Y.~{Wang}, ``{PoBT}: A lightweight consensus algorithm for scalable iot
	business blockchain,'' \emph{IEEE Internet of Things Journal}, vol.~7, no.~3,
	pp. 2343--2355, March 2020.
	
	\bibitem{8595469}
	D.~{Puthal} and S.~P. {Mohanty}, ``Proof of authentication: {IoT}-friendly
	blockchains,'' \emph{IEEE Potentials}, vol.~38, no.~1, pp. 26--29, January
	2019.
	
	\bibitem{8662009}
	D.~{Puthal}, S.~P. {Mohanty}, P.~{Nanda}, E.~{Kougianos}, and G.~{Das},
	``Proof-of-authentication for scalable blockchain in resource-constrained
	distributed systems,'' in \emph{Proc. IEEE International Conference on
		Consumer Electronics (ICCE)}, 2019, pp. 1--5.
	
\end{thebibliography}


\section*{About The Authors}

\vspace{-1.0cm}

\begin{IEEEbiography}
    [{\includegraphics[height=1.25in,keepaspectratio]{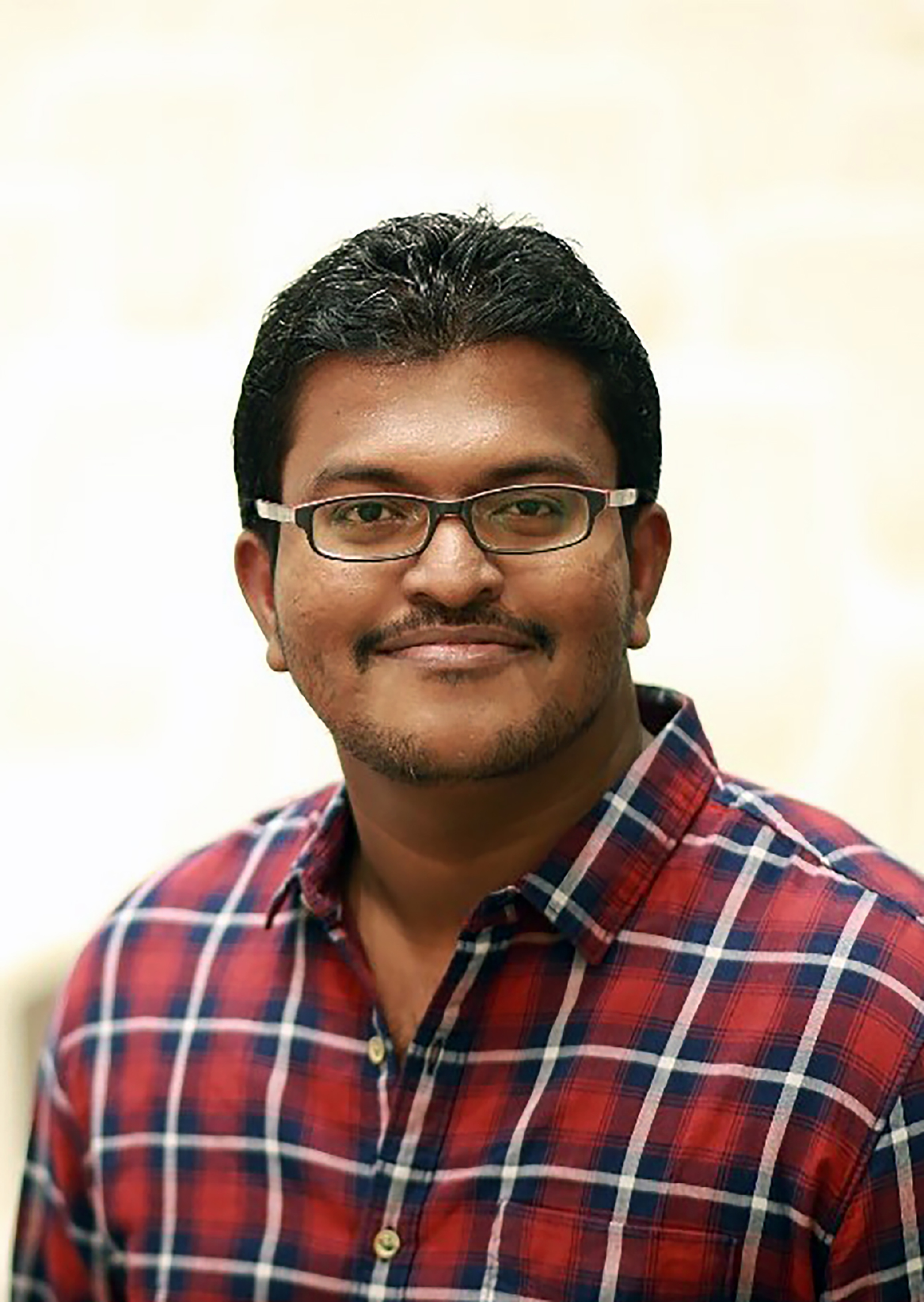}}]
{Sibi C. Sethuraman} (Member, IEEE) received his Ph.D from Anna University in the year 2018 and he is currently working as an Associate Professor in the School of Computer Science and Engineering at Vellore Institute of Technology - Andhra Pradesh (VIT-AP) University since September 2018. He was an Assistant Professor in the Department of Computer Science and Engineering at VIT-AP from May 2018 to September 2018. He is a visiting professor and a member of Artificial Intelligence Lab at University Systems of New Hampshire (KSC campus) in the Department of Computer Science. Further, he is the coordinator for Artificial Intelligence and Robotics (AIR) Center at VIT-AP. He is the lead engineer for the project VISU, an advanced 3D printed humanoid robot developed by VIT-AP. He is an active contributor of open source community. 
He is an active reviewer in many reputed journals of IEEE, Springer, and Elsevier. 
He is a recipient of DST fellowship.
\end{IEEEbiography}

\vspace{-1.0cm}

\begin{IEEEbiography}
[{\includegraphics[height=1.25in,keepaspectratio]{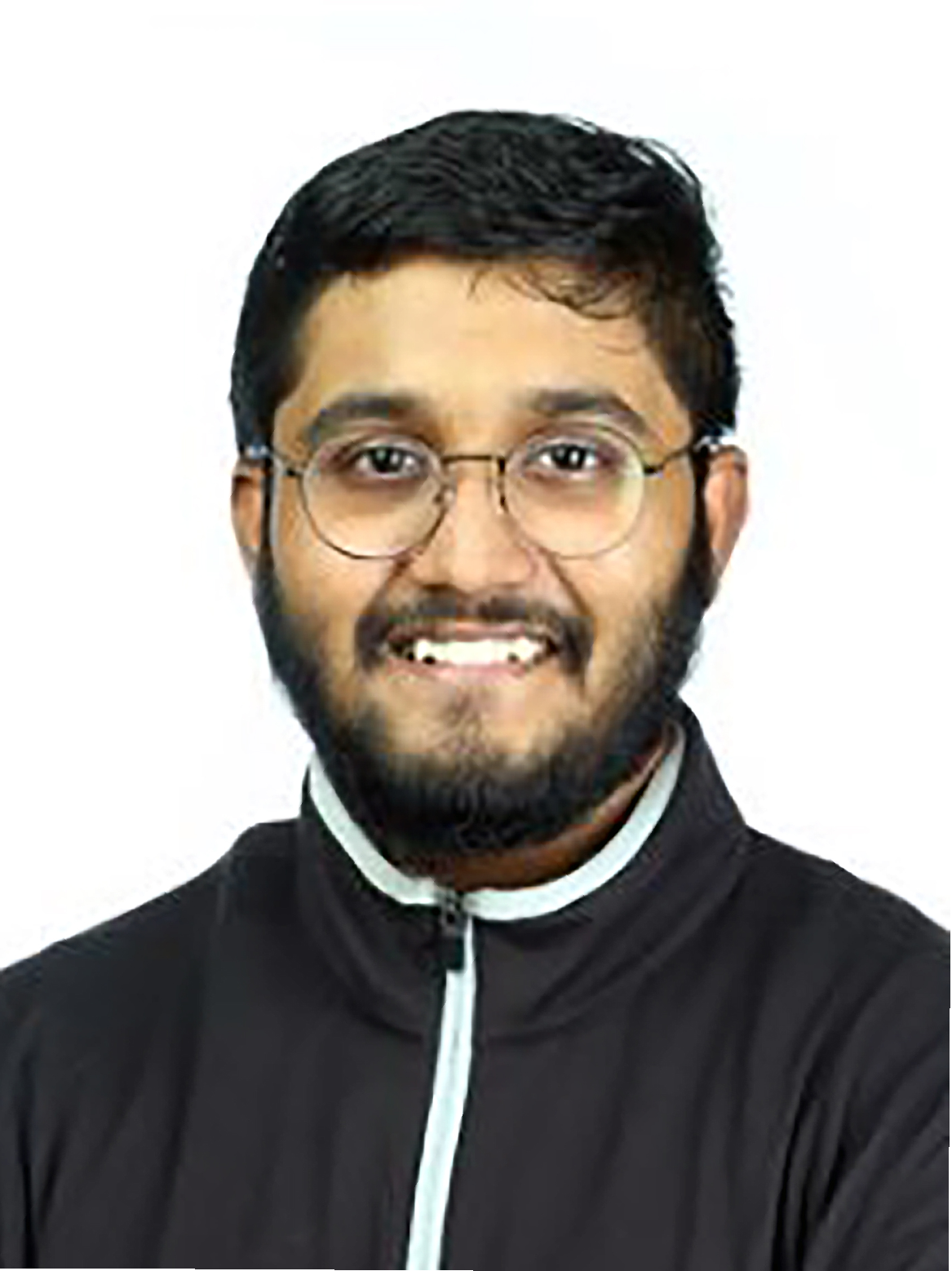}}]
{Pranav Kompally} (Student Member, IEEE) is currently pursuing his Junior year of Bachelor of Technology in Vellore Institute of Technology- Amaravathi (VIT-AP). His areas of interest include Deep Learning Approaches, Internet of Things in Consumer Electronics and Embedded Hardware. He is the Project Lead of ViSU at Artificial Intelligence and Robotics Center, VIT-AP. He is also the
founder and creator of Virtual Navigation System called DPS-StreetView which was Deployed in Delhi Public School, Nacharam. 
\end{IEEEbiography}

\vspace{-1.0cm}

\begin{IEEEbiography}
    [{\includegraphics[height=1.25in,clip,keepaspectratio]{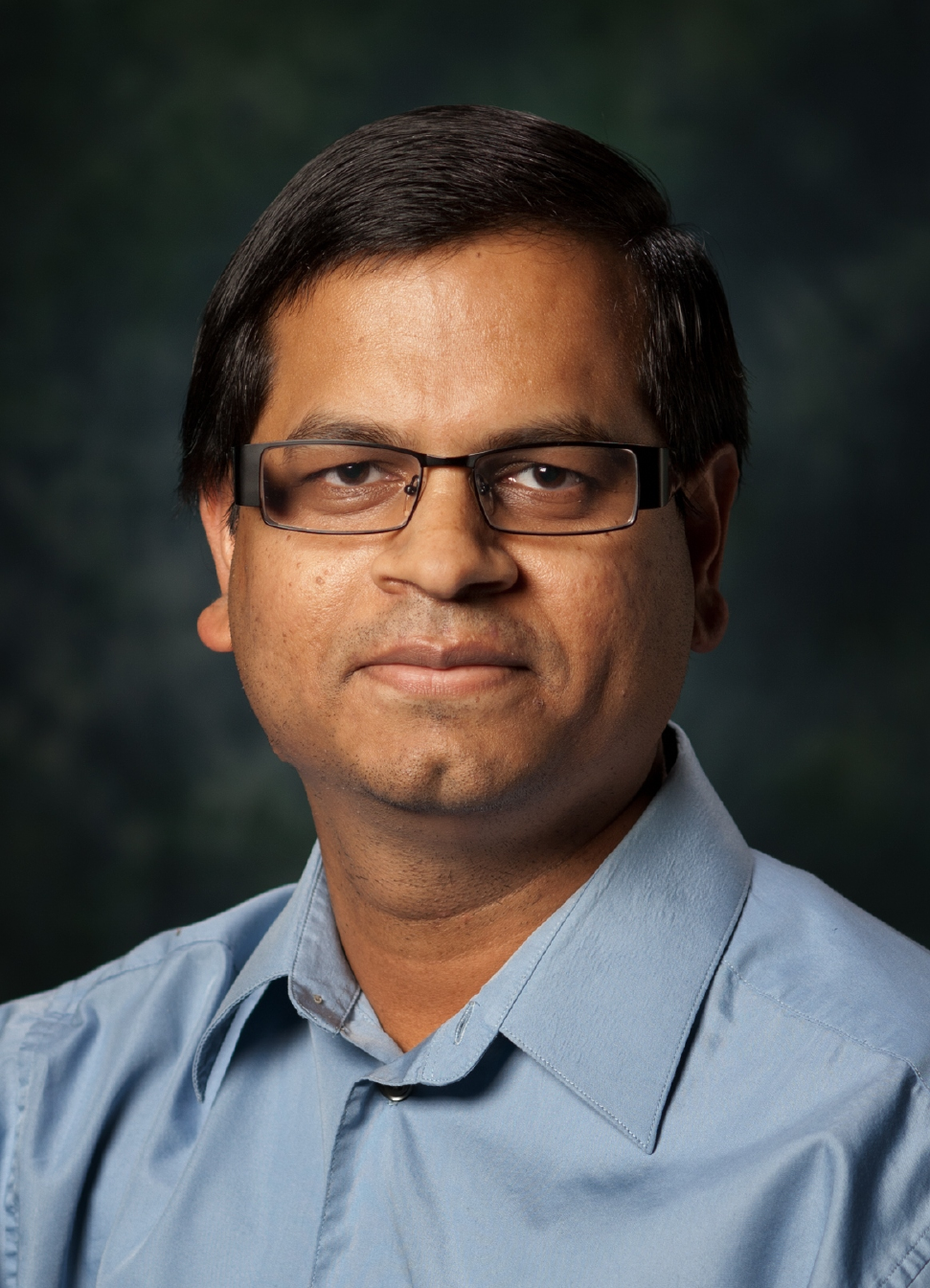}}]
{Saraju P. Mohanty} (Senior Member, IEEE) received the bachelor's degree (Honors) in electrical engineering from the Orissa University of Agriculture and Technology, Bhubaneswar, in 1995, the master's degree in Systems Science and Automation from the Indian Institute of Science, Bengaluru, in 1999, and the Ph.D. degree in Computer Science and Engineering from the University of South Florida, Tampa, in 2003. He is a Professor with the University of North Texas. His research is in ``Smart Electronic Systems'' which has been funded by National Science Foundations (NSF), Semiconductor Research Corporation (SRC), U.S. Air Force, IUSSTF, and Mission Innovation. He has authored 350 research articles, 4 books, and invented 4 U.S. patents. His Google Scholar h-index is 37 and i10-index is 144 with 6200 citations. He is regarded as a visionary researcher on Smart Cities in which his research deals with security and energy aware, and AI/ML-integrated smart components. He introduced the Secure Digital Camera (SDC) in 2004 with built-in security features designed using Hardware-Assisted Security (HAS) or Security by Design (SbD) principle. He is widely credited as the designer for the first digital watermarking chip in 2004 and first the low-power digital watermarking chip in 2006.
He has delivered 9 keynotes and served on 5 panels at various International Conferences. 
He is a recipient of 12 best paper awards, Fulbright Specialist Award in 2020, IEEE Consumer Electronics Society Outstanding Service Award in 2020, the IEEE-CS-TCVLSI Distinguished Leadership Award in 2018, and the PROSE Award for Best Textbook in Physical Sciences and Mathematics category in 2016.  He is the Editor-in-Chief (EiC) of the IEEE Consumer Electronics Magazine (MCE). He has been serving on the editorial board of several peer-reviewed international journals, including IEEE Transactions on Consumer Electronics (TCE), and IEEE Transactions on Big Data (TBD). He has been serving on the Board of Governors (BoG) of the IEEE Consumer Technology Society, and has served as the Chair of Technical Committee on Very Large Scale Integration (TCVLSI), IEEE Computer Society (IEEE-CS) during 2014-2018. He is the founding steering committee chair for the IEEE International Symposium on Smart Electronic Systems (iSES), steering committee vice-chair of the IEEE-CS Symposium on VLSI (ISVLSI), and steering committee vice-chair of the OITS International Conference on Information Technology (ICIT).
He has mentored 2 post-doctoral researchers, and supervised 12 Ph.D. dissertations, 26 M.S. theses, and 10 undergraduate projects.
\end{IEEEbiography}

\vspace{-1.0cm}

\begin{IEEEbiography}
[{\includegraphics[height=1.25in,clip,keepaspectratio]{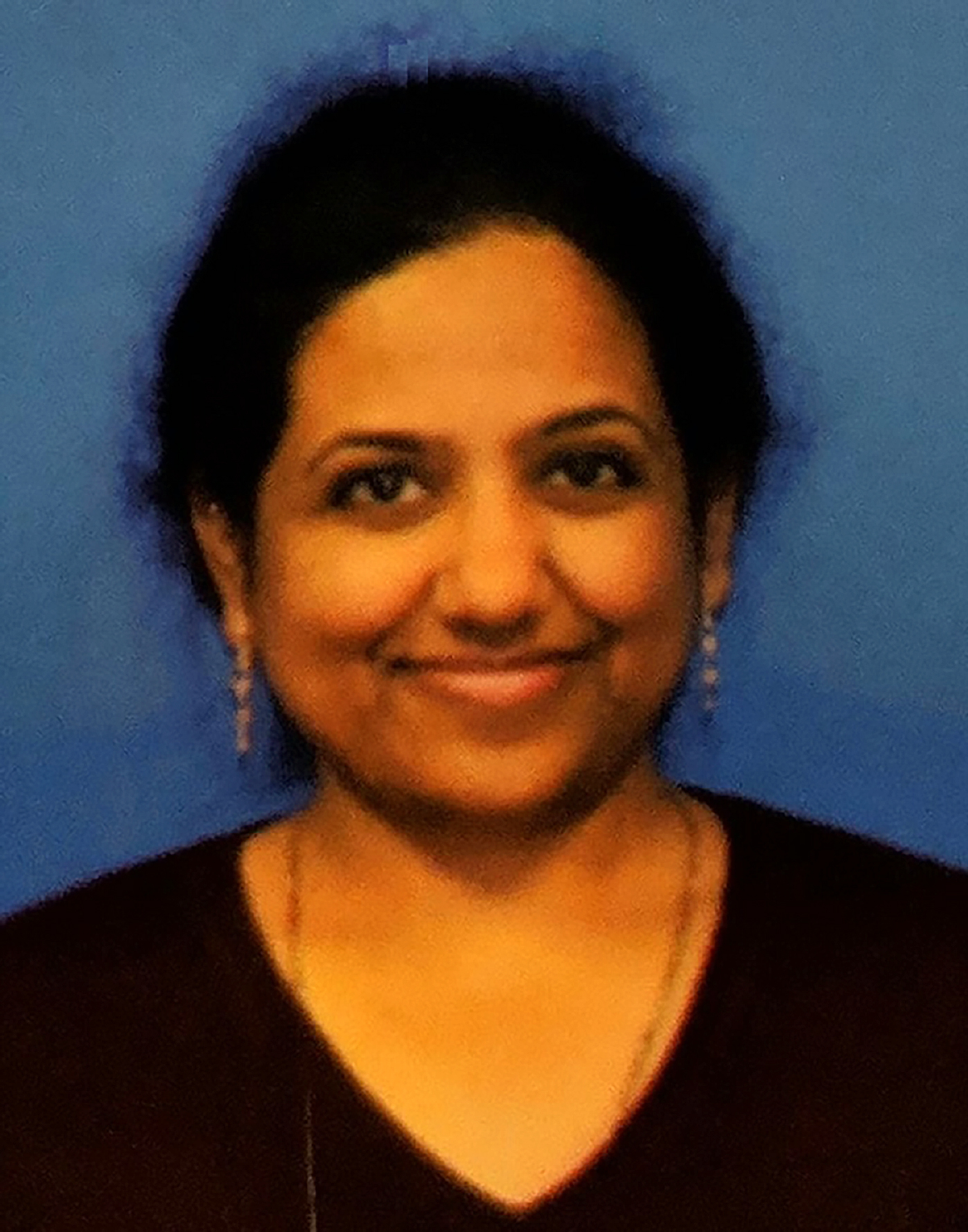}}]
{Uma Choppali} received the bachelor's degree in Science and Education from the Regional Institute of Education (RIE), Bhubaneswar, India, in 1999, the master's degree in Physics from the University of South Florida, Tampa, Florida, USA, in 2004, and the Ph.D. degree in Materials Science and Engineering from the University of North Texas, Denton, in 2006. She is a faculty at the Dallas College - Eastfield Campus, Mesquite, TX, USA. She is an author of a dozen of research articles. Her Google Scholar h-index is 8 and i10-index is 7 with 660 citations. She is a regular reviewer of several International journals and conferences.
\end{IEEEbiography}

\end{document}